\documentclass[journal=jctcce,manuscript=article]{achemso}

\usepackage[english]{babel}

\usepackage[utf8]{inputenc} 
\usepackage{amsmath}
\usepackage{amsfonts}
\usepackage{amssymb}
\usepackage{mathtools}
\usepackage{graphicx}
\usepackage{physics}
\usepackage{tikz}
\usetikzlibrary{fit}
\usetikzlibrary{positioning}
\usetikzlibrary{decorations.pathmorphing, decorations.pathreplacing, decorations.markings, backgrounds}
\usepackage{tikz-network}
\usepackage[justification=centering]{caption,subcaption}
\usepackage[colorlinks=true, allcolors=blue]{hyperref}
\usepackage{makecell}
\usepackage{afterpage}
\usepackage[ruled, vlined, linesnumbered]{algorithm2e}
\usepackage{xcolor, colortbl}
\usepackage{setspace}
\usepackage[version=4]{mhchem}
\usepackage[nohyperlinks, nolist]{acronym}

\graphicspath{{figures/}}
\makeatletter
\def\input@path{{figures/}}
\makeatother

\DeclarePairedDelimiter{\floor}{\lfloor}{\rfloor}

\newcommand{\emptyorb}{\tikz\draw (0,0) circle (1.5mm);}
\newcommand{\fullorb}{\tikz\draw[fill=black] (0,0) circle (1.5mm);}

\pgfdeclaredecoration{simple line}{start}
{
  \state{start}[width = +0pt,
                next state=step]{
    \pgfpathmoveto{\pgfpoint{0pt}{0pt}}
  }
  \state{step}[auto end on length    = 3pt,
               auto corner on length = 3pt,               
               width=+1pt]
  {
    \pgfpathlineto{\pgfpoint{1pt}{0pt}}
  }
  \state{final}
  {}
}

\makeatletter
\def\pgf@stroke@inner@line@if@needed{%
  \ifdim\pgfinnerlinewidth>0pt\relax%
    \let\pgf@temp@save=\pgf@strokecolor@global
    \pgfsys@beginscope%
    {%
      \pgfsetrectcap
      \pgfsys@setlinewidth{\pgfinnerlinewidth}%
      \pgfsetstrokecolor{\pgfinnerstrokecolor}%
      \pgfsyssoftpath@invokecurrentpath%
      \pgfsys@stroke%
    }%
    \pgfsys@endscope%
    \global\let\pgf@strokecolor@global=\pgf@temp@save
  \fi%
}

\title{Wavefunction optimization at the complete basis set limit with Multiwavelets and DMRG}

\addto\captionsenglish{}

\author{Martina Nibbi}
\email{martina.nibbi@tum.de}
\affiliation{Technical University of Munich, School of CIT, Department of Computer Science, Boltzmannstra{\ss}e 3, 85748 Garching, Germany}

\author{Luca Frediani}
\email{luca.frediani@uit.no}
\affiliation{Hylleraas Centre, Department of Chemistry, UiT University of Troms\o{}, The Arctic University of Norway, N-9037 Troms\o{}, Norway}

\author{Evgueni Dinvay}
\email{evgueni.dinvay@uit.no}
\affiliation{Hylleraas Centre, Department of Chemistry, UiT University of Troms\o{}, The Arctic University of Norway, N-9037 Troms\o{}, Norway}

\author{Christian B.~Mendl}
\email{christian.mendl@tum.de}
\affiliation{Technical University of Munich, School of CIT, Department of Computer Science, Boltzmannstra{\ss}e 3, 85748 Garching, Germany}
\affiliation{Technical University of Munich, Institute for Advanced Study, Lichtenbergstra{\ss}e 2a, 85748 Garching, Germany}

\makeatletter
\renewcommand{\@cite}[1]{
{$\!\! ^{(#1)}$}}
\makeatother

\newcommand{\scaling}{\varphi}

\newcommand{\scalingspace}[1]{\ensuremath{{V}^{#1}}}
\newcommand{\waveletspace}[1]{\ensuremath{{W}^{#1}}}
\newcommand{\vampyr}{\texttt{VAMPyR}}

\newcommand{\mrcpp}{\texttt{MRCPP}}
\newcommand{\mrchem}{\texttt{MRChem}}

\newcommand{\madness}{\texttt{M-A-D-N-E-S-S}}

\newcommand{\dalton}{\texttt{Dalton}}
\begin{acronym}
\acro{AO}{atomic orbital}
\acro{AE}{all electron}
\acro{API}{Application Programmer Interface}
\acro{AUS}{Advanced User Support}
\acro{BE}{Binding Energy}
\acrodefplural{BE}[BEs]{Binding Energies}
\acro{BEM}{Boundary Element Method}
\acro{BO}{Born-Oppenheimer}  
\acro{CBS}{complete basis set}
\acro{CC}{Coupled Cluster}
\acro{CI}{Configuration Interaction}
\acro{CTCC}{Centre for Theoretical and Computational Chemistry}
\acro{CoE}{Centre of Excellence}
\acro{DC}{dielectric continuum}  
\acro{DD}{domain decomposition}
\acro{DFT}{density functional theory}  
\acro{DKH}{Douglas-Kroll-Hess}
\acro{DMRG}{density matrix renormalization group}  
\acro{EFP}{effective fragment potential}
\acro{ECP}{effective core potential}
\acro{EU}{European Union}
\acro{FMM}{fast multipole method}
\acro{FCI}{Full Configuration Interaction}
\acro{GGA}{generalized gradient approximation}
\acro{GPE}{Generalized Poisson Equation}
\acro{GTO}{Gaussian Type Orbital}
\acro{HF}{Hartree-Fock}  
\acro{HPC}{high-performance computing}
\acro{Hylleraas}[HC]{Hylleraas Centre for Quantum Molecular Sciences}
\acro{IEF}{Integral Equation Formalism}
\acro{IEFPCM}{Integral Equation Formalism \ac{PCM}}
\acro{IGLO}{individual gauge for localized orbitals}
\acro{IMOM}{Initial Maximum Overlap Method}
\acro{KB}{kinetic balance}
\acro{KS}{Kohn-Sham}
\acro{LAO}{London atomic orbital}
\acro{LAPW}{linearized augmented plane wave}
\acro{LDA}{local density approximation}
\acro{MAD}{mean absolute deviation}
\acro{maxAD}{maximum absolute deviation}
\acro{MM}{molecular mechanics}  
\acro{MCSCF}{multiconfiguration self consistent field}
\acro{MPA}{multiphoton absorption}
\acro{MPO}{matrix product operator}
\acro{MPS}{matrix product state}
\acro{MRA}{multiresolution analysis}
\acro{MSDD}{Minnesota Solvent Descriptor Database}
\acro{MOM}{Maximum Overlap Method}
\acro{MW}{multiwavelet}
\acro{NAO}{numerical atomic orbital}
\acro{NeIC}{nordic e-infrastructure collaboration}
\acro{NO}{natural orbital}
\acro{KAIN}{Krylov-accelerated inexact Newton}
\acro{NMR}{nuclear magnetic resonance}
\acro{NP}{nanoparticle}  
\acro{NS}{Non-Standard}  
\acro{OLED}{organic light emitting diode}
\acro{PAW}{projector augmented wave}
\acro{PBC}{Periodic Boundary Condition}
\acro{PCM}{polarizable continuum model}
\acro{PSP}{pseudopotential}
\acro{PW}{plane wave}
\acro{QC}{quantum chemistry}  
\acro{QM/MM}{quantum mechanics/molecular mechanics}  
\acro{QM}{quantum mechanics}  
\acro{RCN}{Research Council of Norway}
\acro{RDM}{reduced density matrix}
\acrodefplural{RDM}{reduced density matrices}
\acro{RDM1}{1-body reduced density matrix}
\acrodefplural{RDM1}{1-body reduced density matrices}
\acro{RDM2}{2-body reduced density matrix}
\acrodefplural{RDM2}{2-body reduced density matrices}
\acro{RMSD}{root mean square deviation}
\acro{RKB}{restricted kinetic balance}
\acro{SC}{semiconductor}
\acro{SCF}{self-consistent field}
\acro{SCRF}{self-consistent reaction field}
\acro{STSM}{short-term scientific mission}
\acro{SAPT}{symmetry-adapted perturbation theory}
\acro{SERS}{surface-enhanced raman scattering}
\acro{STO}{Slater-Type Orbital}
\acro{WPREL}[WP1]{Work Package 1}
\acro{WPROP}[WP2]{Work Package 2}
\acro{WPAPP}[WP3]{Work Package 3}
\acro{WP}{Work Package}  
\acro{X2C}{exact two-component}
\acro{ZORA}{zero-order relativistic approximation}
\end{acronym}

\begin{document}

\begin{abstract}
The \ac{DMRG} is a powerful numerical technique to solve strongly correlated quantum systems: it deals well with systems which are not dominated by a single configuration (unlike \acl{CC}) and it converges rapidly to the \ac{FCI} limit (unlike truncated \ac{CI} expansions).
In this work, we develop an algorithm integrating \ac{DMRG} within the multiwavelet-based \ac{MRA}.
Unlike fixed basis sets, \aclp{MW} offer an adaptive and hierarchical representation of functions, approaching the complete basis set limit to a specified precision.
As a result, this combined technique leverages the multireference capability of \ac{DMRG} and the \acl{CBS} limit of \ac{MRA} and \aclp{MW}.
More specifically, we adopt a pre-existing Lagrangian optimization algorithm for orbitals represented in the MRA domain and improve its computational efficiency by replacing the original \ac{CI} calculations with \ac{DMRG}. Additionally, we substitute the \aclp{RDM} computation with the direct extraction of energy gradients from the \ac{DMRG} tensors.
We apply our method to small systems such \ce{H2}, \ce{He}, \ce{HeH2}, \ce{BeH2} and \ce{N2}. The results demonstrate that our approach reduces the final energy while keeping the number of orbitals low compared to \ac{FCI} calculations on an atomic orbital basis set.
\end{abstract}

\maketitle
\acresetall
\section{Introduction}
\label{Introduction_section}
The \ac{DMRG} is widely recognized as one of the most powerful numerical techniques for solving strongly correlated quantum systems \cite{Schollwoeck_2011, Chan_2011, White_1999}. This success stems from its favourable computational scaling, polynomial in the number of orbitals for a fixed virtual bond dimension, combined with the ability to efficiently capture strong correlation and converge rapidly toward the \ac{FCI} limit. As a result, DMRG offers much broader applicability than traditional \ac{CI} or \ac{CC} methods, which either converge slowly to \ac{FCI} or are best suited to describe single-determinant dominated systems.
Despite these advantages, the accuracy and efficiency of \ac{DMRG} calculations strongly depend on the choice of the underlying orbital basis. 
Conventional basis sets, such as \acp{GTO} and \acp{PW}, are often limited: the former converges slowly to the \ac{CBS} limit and is inherently non-orthonormal, whereas the latter has trouble describing the cusp region of the nuclei. Larger systems can also require an elevated number of orbitals to reach sufficient precision, which impacts the computational effort of \ac{DMRG}.

In this context, \acp{MW} and \ac{MRA} provide an appealing alternative.
Wavelets and multiwavelets were first constructed in the 80' of last century \cite{Keinert}, with immediate application to signal processing, because they are localized both in real and in Fourier space~\cite{Strang.10.1090/s0273-0979-1993-00390-2}. 
Unlike fixed basis sets, \acp{MW} offer an adaptive and hierarchical representation of functions and allow to reach the \ac{CBS} limit up to any fixed, predefined precision.

Their ability to represent both localized and extended features of molecular wavefunctions makes them particularly well-suited for high-accuracy electronic structure calculations. This was first recognized at the end of the last century by Arias \cite{Arias.10.1103/revmodphys.71.267}. A practical realization of electronic structure calculations using wavelets and multiwavelets was achieved a few years later. The BigDFT code features a wavelet representation of functions and operators, combined with a pseudopotential representation of the core electrons \cite{bigdft}. It is capable of \ac{DFT} calculations of both isolated and periodic systems. The \madness{} code \cite{Harrison_2016} was the first to demonstrate the feasibility of all-electron calculations using \acp{MW}. To achieve an all-electron description, several technical advances are required, which operate in synergy to overcome the obstacles posed by a three-dimensional representation of functions: 
\begin{enumerate}
    \item \ac{SCF} equations for orbital optimization are presented in an integral formulation, enabling the use of Green's function technology\cite{Kalos.10.1103/physrev.128.1791};
    \item adaptive grids, which limit the memory footprint,  are employed to represent functions\cite{Frediani_Fossgaard_Fla_Ruud};
    \item the \ac{NS} form of the operator is required\cite{Beylkin.10.1016/j.acha.2007.08.001} to preserve adaptivity when operators are applied;
    \item operators are represented as a sum of separable terms (generally Gaussians) up to the requested precision\cite{Beylkin.10.1016/j.acha.2005.01.003}, to limit the curse of dimensionality;
    \item localized orbitals are employed\cite{mrchem_2023}, to achieve near-linear scaling with the system size.
\end{enumerate}

All these advances have made it possible to use \acp{MW} for all-electron calculations at the \ac{SCF} level~\cite{Harrison_2016, mrchem_2023, Jensen.10.1021/acs.jpclett.7b00255}. Nevertheless, introducing correlation has for a long time proven to be a formidable task: on the one hand the curse of dimensionality makes it much harder to represent functions in six dimensions, requiring compression techniques such as tensor-train decomposition for the representation of the two-body terms~\cite{Bischoff.10.1063/1.4747538}, or explicit R12 methods to describe the electron-electron cusp effectively\cite{Kong_Bischoff_Valeev2012}. On the other hand, correlated methods are often based on exploiting the (finite) virtual space to go beyond the single-determinant representation. \acp{MW} do not make use of a finite virtual space and one has to devise strategies to obtain such corrections directly, which is possible only for a handful of cases such as MP2\cite{Bischoff.10.1063/1.4820404} or CC2\cite{Kottmann.10.1021/acs.jctc.7b00694,Kottmann.10.1021/acs.jctc.7b00695} and is not of general applicability, because that would require a (finite) set of virtual orbitals. 

In this respect, the recent work from Valeev et al.~\cite{Valeev_2023} provides a general method to deal with correlated wave functions in a \ac{MW} framework.
The correlation problem is handled in a traditional way, namely a \ac{CI}-type optimization, whereas the three-dimensional orbitals are represented with \acp{MW}. The approach relies on employing \acp{NO} which is a crucial step to ensure that orbital energies in the Helmholtz kernel will stay negative.

Integrating \ac{DMRG} into the original optimization algorithm from Valeev et al.~\cite{Valeev_2023} is a natural next step to improve its computational efficiency. Specifically, \ac{DMRG} replaces the original CI calculations, and we introduce an alternative approach to the \acp{RDM} evaluation by extracting energy gradients directly from the \ac{DMRG} tensor network.
This work also represents the first application of \ac{DMRG} within the MRA and \acp{MW} framework.

In this paper, we begin by introducing the \acp{MW} framework as a basis for our method. Building on this, we derive the self-consistent equation for orbital optimization by combining \ac{MRA} with \ac{DMRG}.
While the mathematical derivation follows the same structure as in Valeev et al.~\cite{Valeev_2023}, we present it here for clarity.
This also allows us to highlight certain implementation details that were overlooked in the original work but are crucial for computational efficiency. Numerical benchmarks on small systems show that this approach consistently yields lower energies than standalone \ac{DMRG} or \ac{FCI} calculations on atomic orbital basis sets, owing to the systematic completeness of the \acp{MW} basis. 


\section{Multiresolution analysis and multiwavelets}
\label{Multiresolution_analysis_section}

Alpert's multiwavelets \cite{Alpert1993} are a simple realization of \ac{MW} functions which start from a set of $k+1$ polynomials $\scaling_j(x)$ up to order $k$ in the interval $(0,1)$ and zero otherwise.  They constitute a vector space $\scalingspace{}_0$. Starting from $\scalingspace{}_0$ one can then obtain a ladder of nested spaces $\scalingspace{}_n$, with $n=1,2,3...$ by dilation and translation of the original functions:
\begin{equation}
\label{scaling_dilation_translation}
    \scaling_{j,\ell}^n(x)
    =
    2^{n/2}
    \scaling_j( 2^n x - \ell)
    \; ,
\end{equation}

It can be shown that the sequence of nested spaces $\scalingspace{}_0 \subset \scalingspace{}_1 \subset \scalingspace{}_2 \subset \scalingspace{}_3 \subset ...$ is dense in $L^2$. Wavelet spaces are then constructed by taking the orthogonal complement between two successive scaling spaces, such that $\scalingspace{}_n \oplus \waveletspace{}_n = \scalingspace{}_{n+1}$.

Moreover, by using \acp{MW} one can obtain sparse representations of functions \cite{Alpert_Beylkin_Gines_Vozovoi2002} and certain convolution-type operators \cite{Fann_Beylkin_Harrison_Jordan2004}, such as the Helmholtz operator:
\begin{equation}
\label{eq_helmholtz_operator}
    \hat{G}_{-\lambda} f(\vec{r}) = -2\int \frac{e^{-\sqrt{2\lambda}\vert \vec{r} - \vec{r^\prime}\vert}}{4\pi\vert \vec{r} - \vec{r^\prime}\vert} f(\vec{r^\prime}) d\vec{r^\prime}
    , \quad
    \lambda > 0
    \; .
\end{equation}
This leads to fast and efficient algorithms by maintaining control over the representation error \cite{Frediani_Fossgaard_Fla_Ruud}, which is the distinctive feature of \ac{MW} algorithms.

\begin{figure*}[t]
    \centering
    \subfloat[]{%
    \centering
        \label{fig:scaling_functions}
        \includegraphics[width=0.3\linewidth]{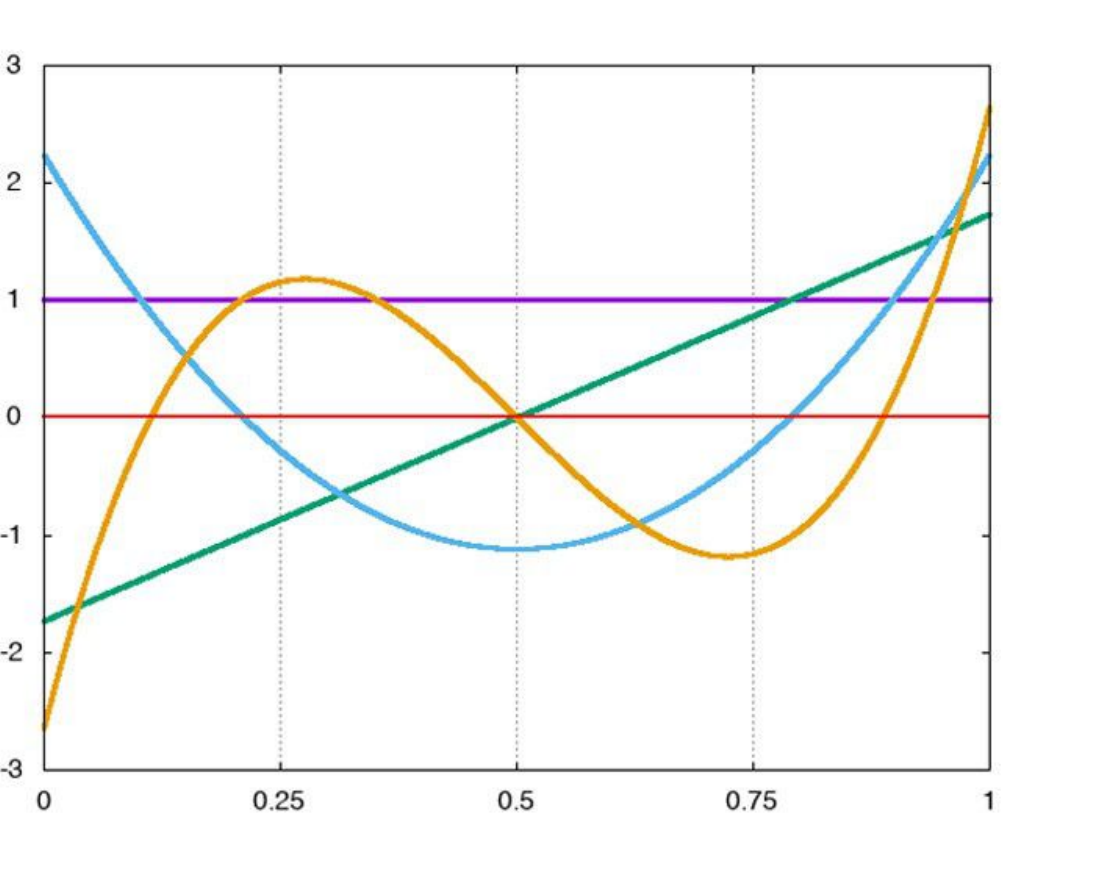}%
    }\hfill
    \subfloat[]{%
        \label{fig:mra_figure}
        \includegraphics[width=0.3\linewidth]{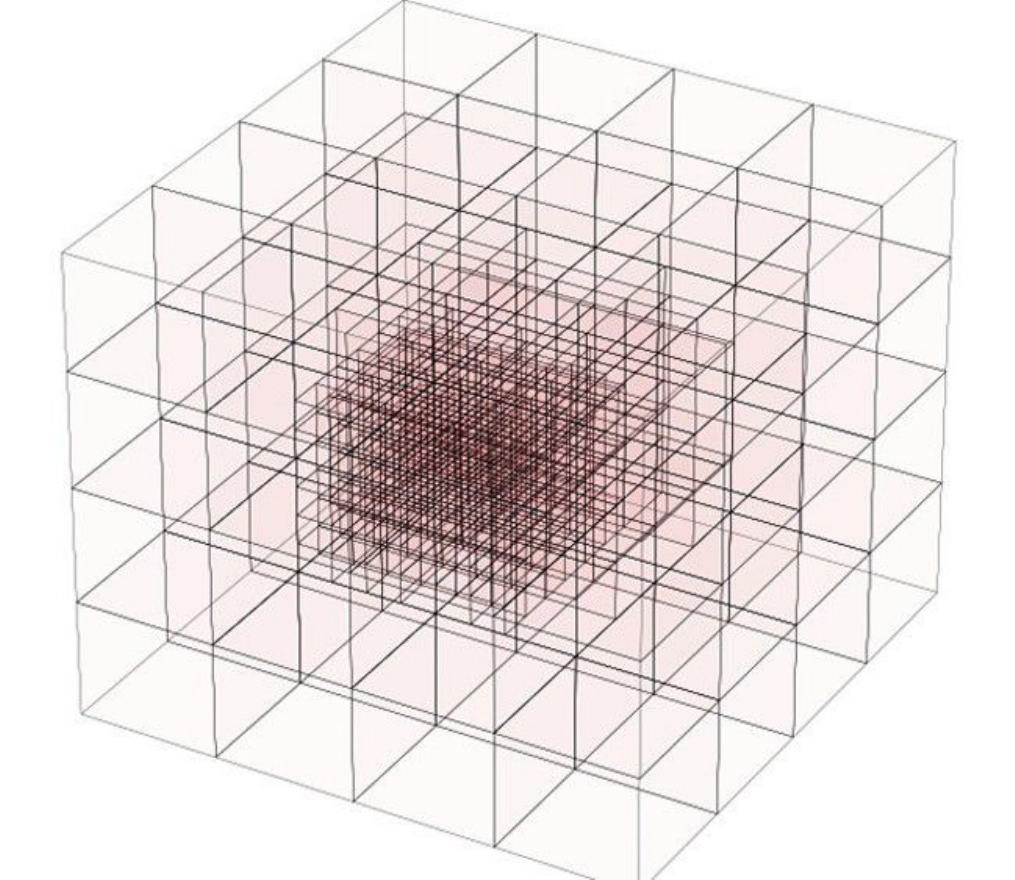}%
    }\hfill
    \subfloat[]{%
        \label{fig:multiwavelets_functions}
        \includegraphics[width=0.3\linewidth]{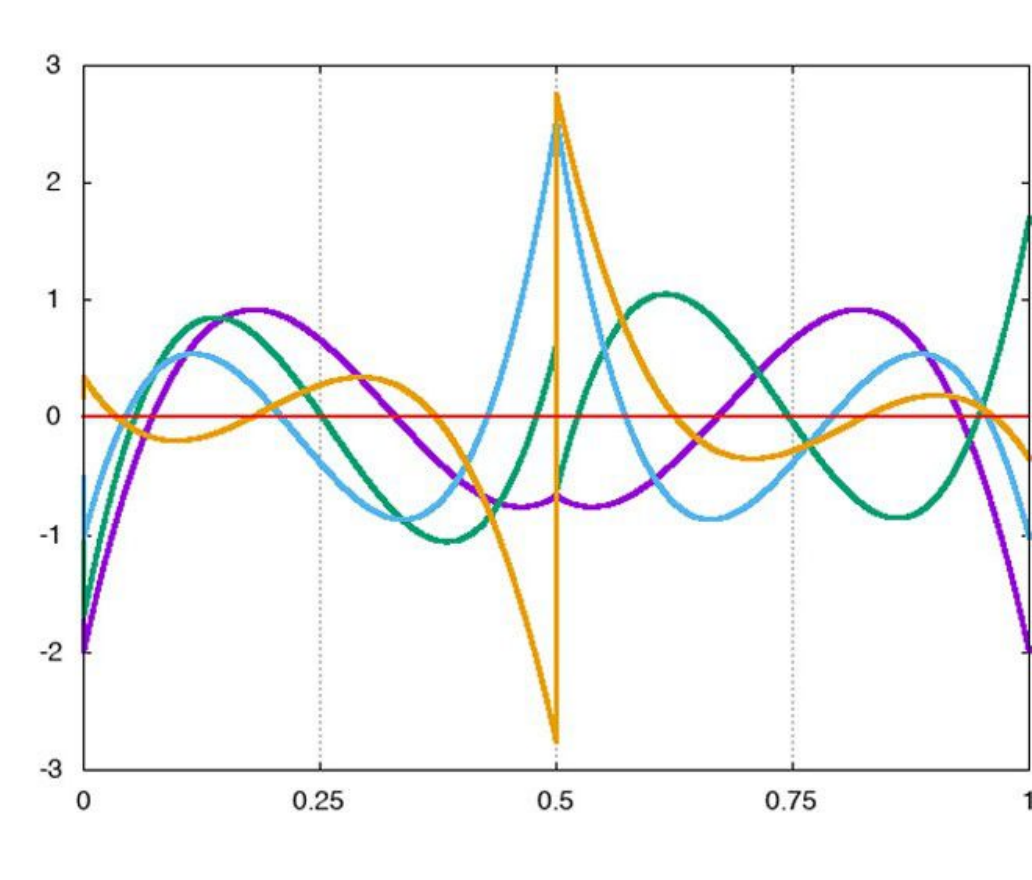}%
    }\hfill
    \caption{(\subref{fig:scaling_functions}) Scaling functions in $V_0$ for $k=3$. (\subref{fig:mra_figure}) Multi-reference grid in $3$D space. (\subref{fig:multiwavelets_functions}) \acp{MW} in $W_0$ for $k=3$. Reproduced from \cite{figures_mra}.}
    \label{fig:scaling_mra_mw}
\end{figure*}

In Fig.~\ref{fig:scaling_mra_mw} we report a graphical representation of the concepts mentioned so far \cite{figures_mra}.
It is beyond the scope of this paper to give a comprehensive exposition of \ac{MW} methods, which have been presented in detail in the literature
\cite{Bischoff2019}.

\section{Combining MRA with DMRG}\label{sec_mra_dmrg}

In this section, we show how to combine MRA with \ac{DMRG} for orbital optimization. Our method takes inspiration from the work of Valeev et al.~\cite{Valeev_2023}, which consists of a multi-reference technique using MRA and \acp{MW}.

Given a finite set of real orbitals $\{\phi_m\}$, we first build the second-quantized chemical Hamiltonian:
\begin{align} \label{eq_def_second_quantization_Hamiltonian}
    \hat{H} &= \sum_{i,j} \sum_{\sigma \in \{\uparrow, \downarrow\}} h_{ij} \, \hat{a}^{\dagger}_{i,\sigma} \hat{a}_{j,\sigma} + \frac{1}{2} \sum_{i,j,k,\ell} \sum_{\sigma,\tau} g_{ijk\ell} \,\hat{a}^{\dagger}_{i,\sigma} \hat{a}^{\dagger}_{j,\tau} \hat{a}_{\ell,\tau} \hat{a}_{k,\sigma} \\
    \label{eq_def_one_two_body_operators}
    &= \hat{h} + \hat{g} \; ,
\end{align}
where $\hat{h}$ and $\hat{g}$ are the one- and two-body operators and with one- and two-body integrals respectively defined as:
\begin{align}
    \label{eq_def_one_body_integral}
    h_{ij} &= \int \phi_i(\vec{r'}) \left[-\frac{\nabla^2}{2} + \sum_\alpha \frac{Z_\alpha}{\vert \vec{r'} - \vec{R}_\alpha \vert}\right] \phi_j(\vec{r'}) \,d\vec{r'} \\
    \label{eq_def_two_body_integral}
    g_{ijk\ell} &= \int \phi_i(\vec{r'}) \phi_j(\vec{r''}) \frac{1}{\vert \vec{r'} - \vec{r''} \vert} \phi_\ell(\vec{r''}) \phi_k(\vec{r'}) \, d\vec{r'} d\vec{r''} \, .
\end{align}
We aim to find the ground-state $\ket{\Phi_{\text{GS}}}$ and the corresponding energy $E = \bra{\Phi_{\text{GS}}}\hat{H}\ket{\Phi_{\text{GS}}}$ through a \ac{DMRG} calculation (see Appendix \ref{appendix_DMRG}).
However, this ground-state energy estimate heavily relies on the choice of orbitals $\{\phi_m\}$ defining $\hat{H}$. As a consequence, in order to get a good energy estimate, we also have to optimize the underlying orbitals.

Techniques involving atomic orbitals rely on a fixed basis, while the present method optimizes the orbitals at the \ac{CBS} limit.
In particular, we want to optimize the orbitals while preserving their orthogonality. This translates to a constraint optimization with Lagrangian:
\begin{equation}\label{eq_def_Lagrangian}
    \mathcal{L} = E - \sum_{i,j}\varepsilon_{ij}(s_{ij} - \delta_{ij}) \; ,
\end{equation}
where $s$ is the orbitals' overlap matrix and $\varepsilon$ the Lagrange multipliers' matrix.

Following Valeev et. al.~\cite{Valeev_2023}, we differentiate $\mathcal{L}$ with respect to variations of the orbitals $\phi_m$, we get:
\begin{gather}\label{eq_def_Lagrangian_zero}
    \ket{\frac{\partial \mathcal{L}}{\partial \phi_m}}  = 0 \\
    \label{eq_def_Lagrangian_gradient}
    \frac{\partial E}{\partial \phi_m} - \sum_{ij} \varepsilon_{ij} \frac{\partial s_{ij}}{\partial \phi_m} = 0 \; .
\end{gather}
The first term in Eq.~\eqref{eq_def_Lagrangian_gradient} can be expanded by using the chain rule:
\begin{equation} \label{eq_chain_rule}
    \frac{\partial E}{\partial \phi_m} = \sum_{ij} \frac{\partial h_{ij}}{\partial \phi_m}\frac{\partial E}{\partial h_{ij}} + \sum_{ijk\ell}\frac{\partial g_{ijk\ell}}{\partial \phi_m}\frac{\partial E}{\partial g_{ijk\ell}} \; ,
\end{equation}
where $h_{ij}$ and $g_{ijk\ell}$ have been defined respectively in Eq.~\eqref{eq_def_one_body_integral} and Eq.~\eqref{eq_def_two_body_integral}.

This notation is particularly convenient because the terms $\partial E/\partial h_{ij}$ and $\partial E/\partial g_{ijk\ell}$ can be easily derived by ``cutting holes" in the \ac{MPO} used for the \ac{DMRG} calculation, as shown in Appendix \ref{appendix_DMRG}.
It is worth remarking that instead of the energy gradients, Valeev et al.~\cite{Valeev_2023} evaluated the one- and two-body \acp{RDM}.
While calculated in a different way, it can easily be proven that these quantities are equivalent:
\begin{subequations}
\label{eq:energy_gradients_rdms}
\begin{gather}
    \frac{\partial E}{\partial h_{ij}} = \bra{\Phi_{\text{GS}}} \hat{E}_{ij} \ket{\Phi_{\text{GS}}} = \gamma_{ij} \\
    \frac{\partial E}{\partial g_{ijk\ell}} = \bra{\Phi_{\text{GS}}} \hat{e}_{ijk\ell} \ket{\Phi_{\text{GS}}} = \gamma_{ijk\ell} \; .
\end{gather}
\end{subequations}
 where the spin-free 1-electron and 2-electron excitation operators are defined as 
 $\hat{E}_{ij} = \sum_{\sigma} a^{\dagger}_{i,\sigma} a_{j,\sigma}$ and 
 $\hat{e}_{ijk\ell} = \hat{E}_{ik}\hat{E}_{j\ell}- \delta_{jk}\hat{E}_{i\ell} = \sum_{\sigma,\tau} a^{\dagger}_{i,\sigma} a^\dagger_{j,\tau} a_{\ell,\tau} a_{k,\sigma}$, following the conventions used by Helgaker, Jørgensen and Olsen~\cite{10.1002/9781119019572}.

As a result, our mathematical derivation is also equivalent.
However, we chose to present the same proof again as this allows us to focus more on the implementation details, which were sometimes missing from the original work and are fundamental when taking efficiency into account.

The first step consists of the analytic calculation of $\partial h_{ij}/\partial \phi_m$ and $\partial g_{ijk\ell}/\partial \phi_m$.
More specifically, for the one-body term, we get:
\begin{equation}
    \frac{\partial h_{ij}}{\partial \phi_m} =  \left(\delta_{im} \hat{h} \ket{\phi_j} + \delta_{jm}\hat{h}\ket{\phi_i}\right) \; .
\end{equation}
The first term of Eq.~\eqref{eq_chain_rule} looks then like:
\begin{align}
    \sum_{ij}\frac{\partial h_{ij}}{\partial \phi_m}\frac{\partial E}{\partial h_{ij}} &= 
    \sum_j \left[\frac{\partial E}{\partial h_{mj}} + \frac{\partial E}{\partial h_{jm}}\right] \hat{h} \ket{\phi_j} \nonumber \\
    &= 2\sum_j \partial E^{(1)}_{mj} \,\hat{h} \ket{\phi_j} \; ,
\end{align}
where we defined the symmetrized version of the one-body gradient as:
\begin{equation}
    \partial E^{(1)} =  \frac{1}{2}\left[\frac{\partial E}{\partial h} +  \left(\frac{\partial E}{\partial h}\right)^T \right] \; .
\end{equation}

Similarly, for what concerns the two-body term, we derive:
\begin{align}
    \frac{\partial g_{ijk\ell}}{\partial \phi_m}
    &= \delta_{im} \hat{g}_{j\ell}\ket{\phi_k} + \delta_{jm} \hat{g}_{ik} \ket{\phi_\ell} \nonumber \\
    &+ \delta_{km} \hat{g}_{j\ell} \ket{\phi_i} + \delta_{\ell m} \hat{g}_{ik} \ket{\phi_j}  ,
\end{align}
where the operator $\hat{g}_{jl}$ is defined as:
\begin{equation}\label{eq_def_g_jl}
    \hat{g}_{j\ell} = \int \phi_j(\vec{r'}) \frac{1}{\vert \vec{r} - \vec{r'} \vert} \phi_\ell(\vec{r'})  \, d\vec{r'} \; .
\end{equation}
The second term of Eq.~\eqref{eq_chain_rule} can then be written as:
\begin{align}
    \sum_{ijk\ell}\frac{\partial g_{ijk\ell}}{\partial \phi_m}\frac{\partial E}{\partial g_{ijk\ell}} 
    &= \sum_{jk\ell} \left[
        \frac{\partial E}{\partial g_{mjk\ell}} + \frac{\partial E}{\partial g_{jm\ell k}}
        + \frac{\partial E}{\partial g_{kjm\ell}} + \frac{\partial E}{\partial g_{jk\ell m}} \right] \hat{g}_{j\ell} \ket{\phi_k}
        \nonumber \\
    &= 4\sum_{jk\ell} \partial E^{(2)}_{mjk\ell} \, \hat{g}_{j\ell} \ket{\phi_k} \; ,
\end{align}
where the symmetrized version of the two-body gradient is defined as:
\begin{equation}
    \partial E^{(2)} = \frac{1}{4}\bigg[\frac{\partial E}{\partial g} \; + \; \left(\frac{\partial E}{\partial g}\right)^{T_{1,0,3,2}} + \left(\frac{\partial E}{\partial g}\right)^{T_{2,1,0,3}} + \left(\frac{\partial E}{\partial g}\right)^{T_{1,2,3,0}}\bigg] 
\end{equation}
and the transposition $T_{i,j,k,\ell}$ switches the indices $0\rightarrow i$, $1\rightarrow j$, $2\rightarrow k$ and $3\rightarrow \ell$. 

Finally, we apply the same derivation to the multipliers' term in Eq.~\eqref{eq_def_Lagrangian_gradient}:
\begin{align}\label{eq_lagrange_multipliers}
    \sum_{ij}\varepsilon_{ij} \frac{\partial s_{ij}}{\partial \phi_m} &= 
    \sum_{ij} \varepsilon_{ij} \, \frac{\partial}{\partial \phi_m} \int \phi_i(\vec{r'})\phi_j(\vec{r'}) \,d\vec{r'} \nonumber\\
    &= \sum_{ij}\left(\delta_{im}\ket{\phi_j} + \delta_{mj}\ket{\phi_i}\right)\varepsilon_{ij} \nonumber\\
    &= \sum_j \left(\varepsilon_{jm} + \varepsilon_{jm}\right)\ket{\phi_j} \nonumber \\
    &= 2\sum_j \bar{\varepsilon}_{mj} \ket{\phi_j} \; ,
\end{align}
with the symmetrized multipliers matrix defined as:
\begin{equation}
    \bar{\varepsilon} = \frac{1}{2}\left[\varepsilon + \varepsilon^T\right] \; .
\end{equation}

As a result, we get the following expression for the vanishing gradients of the Lagrangian w.r.t. the orbitals:
\begin{equation}\label{eq_Lagrangian_derivate_equation}\hspace{-0.21cm}
    \sum_j \left(\partial E^{(1)}_{mj}\hat{h} - \bar{\varepsilon}_{mj}\right)  \ket{\phi_j} + 2\sum_{jk\ell} \partial E^{(2)}_{mjk\ell} \, \hat{g}_{j\ell} \ket{\phi_k}=0  \,.
\end{equation}

We can derive the Lagrangian multipliers by projecting Eq.~\eqref{eq_Lagrangian_derivate_equation} on the orthonormal orbitals:
\begin{equation}
    \braket{\phi_n}{\frac{\partial \mathcal{L}}{\partial \phi_m}} = 0 \; ,
\end{equation}
from which we get:
\begin{equation}\label{eq_lagrange_multipliers_derivation}
    \bar{\varepsilon}_{mn} = \sum_j \partial E^{(1)}_{mj} \, h_{nj} + 2\sum_{jk\ell} \partial E^{(2)}_{mjk\ell} \, g_{njk\ell} \; .
\end{equation}

To simplify the equations, Valeev et al.~\cite{Valeev_2023} used \acp{NO}, which diagonalize the one-body \ac{RDM}. 
We achieve the same outcome, by diagonalizing the one-body energy gradient.
Given then the basis-change matrix $U$ such that:
\begin{equation}\label{eq_diagonalization_one_body_grad}
    \partial E^{(1)} = U^{\dagger} \Lambda U \; ,
\end{equation}
where $\Lambda$ is diagonal, we can derive Eq.~\eqref{eq_Lagrangian_derivate_equation} in the new basis for every $m$:
\begin{equation}
    \Lambda_m\hat{h}\ket{\phi^\prime_m} + 
    2\sum_{ijk\ell} U_{mi}\, \partial E^{(2)}_{ijk\ell} \, \hat{g}_{j\ell}\ket{\phi_k} -\sum_{j} \varepsilon^{\prime}_{mj}\ket{\phi^\prime_j} =0 \; ,
\end{equation}
where:
\begin{equation}
    \ket{\phi^\prime_i} = \sum_j U_{ij}\ket{\phi_j}
\end{equation}
and:
\begin{equation}
    \varepsilon^\prime = U \bar{\varepsilon}\, U^{\dagger} \;  .
\end{equation}
The two-body term is not rotated because the summation is carried out over the whole set. We notice in this respect that this flexibility could be exploited to compute the two-body term on a localized basis to achieve a reduced scaling of this time-consuming step.

Finally, we obtain the self-consistent equation:
\begin{equation}\label{eq_working_equation_differential}
    -\left(\hat{d} - \frac{\varepsilon^{\prime}_{mm}}{\Lambda_m}\right)\ket{\phi^\prime_m} =
    \hat{v}\ket{\phi^\prime_m} + \frac{1}{\Lambda_m}\left(2\sum_{ijk\ell}  U_{mi} \partial E^{(2)}_{ijk\ell} \, \hat{g}_{j\ell}\ket{\phi_k} - \sum_{j\neq m} \varepsilon^{\prime}_{mj}\ket{\phi^\prime_j}\right) \; ,
\end{equation}
where the one-body operator $\hat{h} = \hat{d} + \hat{v}$ has been split into its kinetic and potential contributions.

The orbital update equation is then derived by inverting Eq.~\eqref{eq_working_equation_differential}:
\begin{equation}\label{eq_optimization}
    \ket{\phi^\prime_m} = -\hat{G}_{\frac{\varepsilon^{\prime}_{mm}}{\Lambda_m}} \left(\hat{v}\ket{\phi^\prime_m} + \frac{1}{\Lambda_m}\left(2\sum_{ijk\ell} U_{mi} \partial E^{(2)}_{ijk\ell} \, \hat{g}_{j\ell}\ket{\phi_k} - \sum_{j\neq m} \varepsilon^{\prime}_{mj}\ket{\phi^\prime_j}\right)\right) \; ,
\end{equation}
where the Helmholtz operator $\hat{G}_{-\lambda}$ was defined in Eq.~\eqref{eq_helmholtz_operator}.

It is worth remarking that the inversion of Eq.~\eqref{eq_working_equation_differential} is practical only if the coefficients $\varepsilon^\prime_{mm}/\Lambda_m$ are negative.
By rewriting Eq.~\eqref{eq_lagrange_multipliers_derivation} in the new orbitals' basis, we get:
\begin{equation}\label{eq_sign_Helmholtz_coefficients}
    \frac{\varepsilon^\prime_{mm}}{\Lambda_m} = h^{\prime}_{mm} + 2\sum_{jk\ell}\frac{ U_{mi}U_{mn}\partial E^{(2)}_{ijk\ell}\,g_{njk\ell}}{\Lambda_m} \; ,
\end{equation}
where $h^{\prime}_{mm}$ is the one-body integral in the new orbitals' basis and $U$ is the base change matrix from Eq.~\eqref{eq_diagonalization_one_body_grad}.
Valeev et al.~\cite{Valeev_2023} came to the same definition of the Helmholtz kernel's coefficients and they argued that, at least empirically, they always tend to be negative since $h^\prime_{mm}$ is always negative and the second term in equation Eq.~\eqref{eq_sign_Helmholtz_coefficients} is small in absolute value.
While this is not a mathematical proof, we have also experienced it to hold for most of the numerical tests we have performed.

\begin{figure*}[t]
    \centering
    \includegraphics[width=\linewidth]{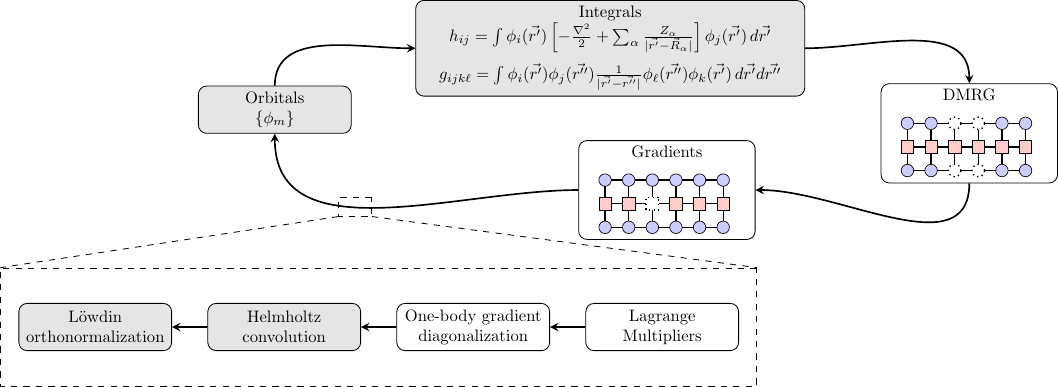}
    \caption{MRA+\ac{DMRG} flowchart. The steps colored in gray belong to the ``MRA domain". The white blocks are orbital-agnostic.}
    \label{fig:algorithm}
\end{figure*}

In conclusion, we were able to derive the same self-consistent working equation as Valeev et al.~\cite{Valeev_2023}.
However, while they evaluated the \acp{RDM} through a Heat-Bath \ac{CI} calculation \cite{heat_bath_2016}, we can obtain the same quantities at a much lower computational effort through a \ac{DMRG} calculation, which also provides a precise energy estimate through the entire optimization process.
The orbitals' optimization algorithm combining MRA and \ac{DMRG} is then summarized in Algorithm~\ref{alg:mra_dmrg} and in the flowchart in Fig.~\ref{fig:algorithm}, where we have also highlighted which steps are based on the MRA representation and which can be considered orbital-agnostic. This constitutes the first algorithm ever leveraging \ac{DMRG} in a \ac{MRA} and \acp{MW} framework.
Note also that, unlike the original work, we perform a L{\"o}wdin orthonormalization of the orbitals \cite{loewdin_1950} as a final step of each iteration in order to smoothen the convergence.

\begin{algorithm}[t]
    \setstretch{1.05}
    \SetKwInOut{Proc}{Procedure}
    \KwIn{Starting set of orthonormal orbitals $\{\phi_m\}$, MRA precision $p$, energy convergence threshold $\delta$}
    \KwOut{Optimal orbitals $\{\phi_m\}$ and gr<fuound state energy}
    \Proc{}
    $ $ \While{$\Delta E > \delta $}{
         Calculate $h_{ij}$ and $g_{ijk\ell}$ from Eq.~\eqref{eq_def_one_body_integral} and~\eqref{eq_def_two_body_integral} \\
         Construct \ac{MPO} of $\hat{H}$ \\ 
         Run \ac{DMRG} and find $E$ and $\ket{\Phi_{\text{GS}}}$ \\
         Extract $\partial E/\partial h_{ij}$ and $\partial E/ \partial g_{ijk\ell}$ \\
         Compute $\varepsilon_{ij}$ from Eq.~\eqref{eq_lagrange_multipliers_derivation} \\
         Diagonalize $\partial E/\partial h_{ij}$ \\
         Use Eq.~\eqref{eq_optimization} to find the new set of orbitals \\
         L\"{o}wdin orthonormalization \cite{loewdin_1950}
    }
    \caption{MRA+DMRG algorithm}\label{alg:mra_dmrg}
\end{algorithm}

\section{Numerical results} 
\label{sec_numerical_results}

We have tested our algorithm on a few examples, where correlation effects are known to play an important role\cite{10.1063/5.0029339, Ganoe.10.1039/d4fd00066h} despite their small size, and we report a comparison with the Hartree Fock method plus the pure application of \ac{DMRG} and \ac{FCI} acting on an atomic orbital basis set.
All MRA calculations have precision $p=10^{-5}$ and polynomial order $k=9$. Moreover, we have fixed the energy convergence threshold $\delta=10^{-5} \text{ Ha}$.
All the experiments were conducted on a workstation computer equipped with an AMD EPYC 7402 CPU 2.80GHz, 24 cores, and 256 GB RAM.

For \ac{MRA} and \acp{MW} representation, we have relied on the Python software \vampyr{}~\cite{vampyr_2024}, which is a Python interface to the underlying C++ mathematical library \mrcpp{}~\cite{bast_2023_7967323}.  At the same time, \ac{DMRG} and gradient extraction are implemented in the \texttt{chemtensor}~\cite{chemtensor} library and are described in Appendix~\ref{appendix_DMRG}. Neither library is currently optimized for a \ac{HPC} framework, which prevented us from considering larger systems.
Therefore, we only provide results for up to 15 orbitals, while we leave the task of optimizing the code for future work.
The \ac{HF} and \ac{FCI} energies have been calculated with \dalton{}~\cite{daltonpaper} with atomic basis sets up to 28 orbitals.

\begin{figure*}
    \centering
    \subfloat{%
        \label{fig:H2_results}
        \subfloat{%
            \includegraphics[width=0.48\linewidth]{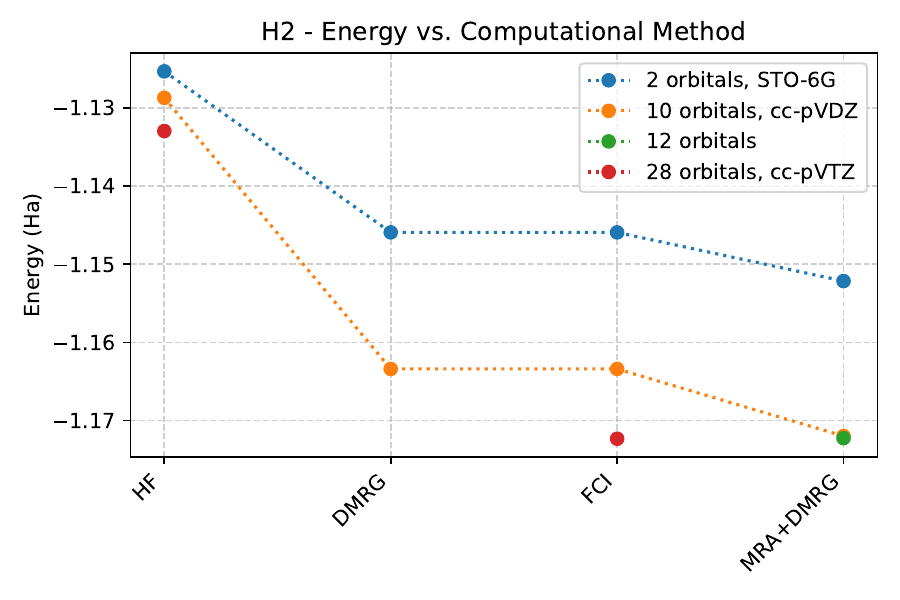}%
        }
        \subfloat{%
            \begin{minipage}{0.5\linewidth}
                \subfloat{
                    \hspace{3.5cm}
                    \begin{tikzpicture}[scale=0.6, transform shape]
    \definecolor{hydrogen}{RGB}{173,216,230} 

    \draw[thick] (0,0) -- (1.5,0);
    
    \shade[ball color=hydrogen] (0,0) circle (0.5);
    \node at (0,0) {\textbf{H}};
    
    \shade[ball color=hydrogen] (1.5,0) circle (0.5);
    \node at (1.5,0) {\textbf{H}};

    \draw[<->] (0,-0.8) -- (1.5,-0.8) node[midway, below] {1.4 a.u.};
\end{tikzpicture}
                    \vspace{0.3cm}
                }\\
                \subfloat{
                    \def\arraystretch{1.5}
                    \resizebox{\linewidth}{!}{%
                    \begin{tabular}{|c|c|c|c|c|}
                        \hline
                        Starting basis & \ac{HF} & \ac{DMRG} & \ac{FCI} & MRA+\ac{DMRG} \\
                        \hline\hline
                        STO-6G (2 orbitals) & -1.12532436 & -1.14592924 & -1.14592924 & -1.15215850 \\
                        cc-pVDZ (10 orbitals) & -1.12870945 & -1.16339873 & -1.16339873 & -1.17199281 \\
                        (12 orbitals) & -- & --  & -- & -1.17226155 \\
                        cc-pVTZ (28 orbitals) & -1.13296053 & -- & -1.17233459 & -- \\
                        \hline
                    \end{tabular}}
                }
            \end{minipage}
            \vspace{1.26cm}
        }
    }\\
    \subfloat{
        \label{fig:He_results}
        \subfloat{%
            \includegraphics[width=0.48\linewidth]{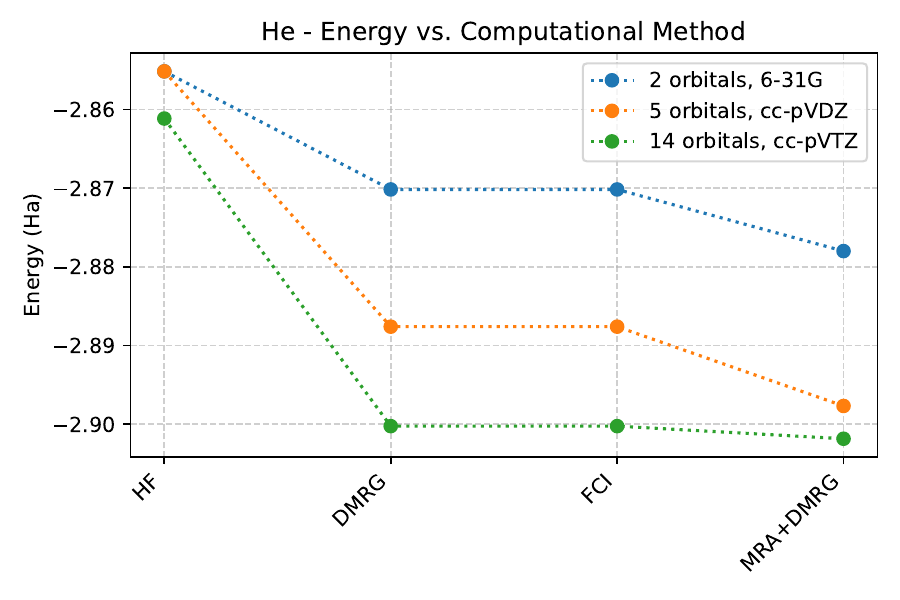}%
        }
        \subfloat{%
            \begin{minipage}{0.5\linewidth}
                \subfloat{
                    \hspace{3.9cm}
                    \begin{tikzpicture}[scale=0.6, transform shape]
    \definecolor{helium}{RGB}{255,200,100} 

    \shade[ball color=helium] (0,0) circle (0.6);
    \node at (0,0) {\textbf{He}};
    
\end{tikzpicture}
                    \vspace{0.8cm}
                }\\
                \subfloat{
                    \def\arraystretch{1.5}
                     \resizebox{\linewidth}{!}{%
                    \begin{tabular}{|c|c|c|c|c|}
                        \hline
                        Starting basis & HF & \ac{DMRG} & \ac{FCI} & MRA+\ac{DMRG} \\
                        \hline\hline
                        6-31G (2 orbitals) & -2.85516043 & -2.87016214 & -2.87016214 & -2.87799355 \\
                        cc-pVDZ (5 orbitals) & -2.85516048 & -2.88759483 & -2.88759483 & -2.89767133 \\
                        cc-pVTZ (14 orbitals) & -2.86115334 & -2.90023217 & -2.90023217 & -2.90183514 \\
                        \hline
                    \end{tabular}}
                }
            \end{minipage}
            \vspace{1.26cm}
        }
    }\\
    \subfloat{%
        \label{fig:HeH2_results}
        \subfloat{%
            \includegraphics[width=0.48\linewidth]{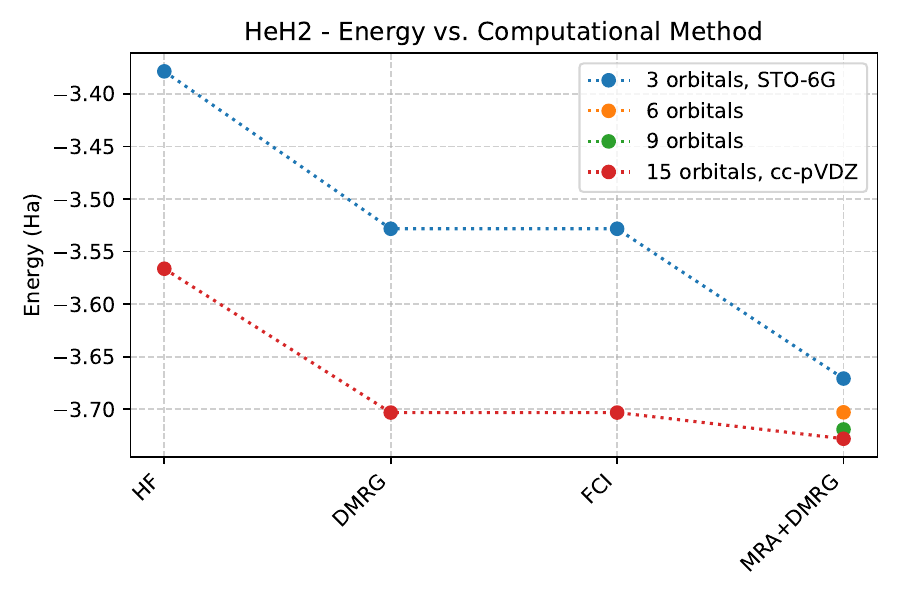}%
        }
        \subfloat{%
            \begin{minipage}{0.5\linewidth}
                \subfloat{
                    \hspace{2.9cm}
                    \begin{tikzpicture}[scale=0.6, transform shape]
    \definecolor{helium}{RGB}{255,200,100} 
    \definecolor{hydrogen}{RGB}{173,216,230} 

    \coordinate (H1) at (-1.3,-0.2);
    \coordinate (He) at (0,1);
    \coordinate (H2) at (1.3,-0.2);

    \draw[thick] (H1) -- (He);
    \draw[thick] (H2) -- (He);
    
    \shade[ball color=hydrogen] (H1) circle (0.5);
    \node at (H1) {\textbf{H}};
    
    \shade[ball color=helium] (He) circle (0.6);
    \node at (He) {\textbf{He}};
    
    \shade[ball color=hydrogen] (H2) circle (0.5);
    \node at (H2) {\textbf{H}};

    \draw[<->] (-1.8, 0.5) -- (-0.5,1.7) node[midway, sloped, above] {1.81 a.u.};

    \draw[] ($ (H1)!0.5!(He) $) arc[start angle=229, end angle=311, radius=1cm];
    \node at (0,-0.2) {104.5°};
\end{tikzpicture}
                    \vspace{0.2cm}
                }\\
                \subfloat{
                    \def\arraystretch{1.5}
                     \resizebox{\linewidth}{!}{%
                    \begin{tabular}{|c|c|c|c|c|}
                        \hline
                        Starting basis & HF & \ac{DMRG} & \ac{FCI} & \ac{MRA}+\ac{DMRG} \\
                        \hline\hline
                        STO-6G (3 orbitals) & -3.37844171 & -3.52823753 & -3.52823753 & -3.67081294 \\
                        (6 orbitals) & -- & -- & -- & -3.70294669 \\
                        (9 orbitals) & -- & -- & -- & -3.71924227 \\
                        cc-pVDZ (15 orbitals) & -3.56627773 & -3.70317251 & -3.70317291 & -3.72808557 \\
                        \hline
                    \end{tabular}}
                }
            \end{minipage}
            \vspace{1.26cm}
        }
    }\\
    \subfloat{%
        \label{fig:BeH2_results}
        \subfloat{%
            \includegraphics[width=0.48\linewidth]{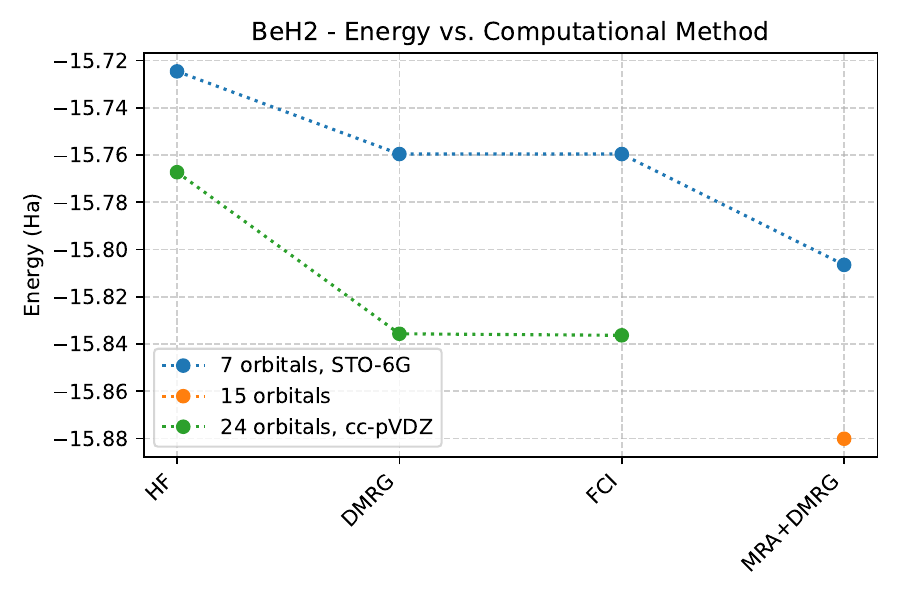}%
        }
        \subfloat{%
             \begin{minipage}{0.5\linewidth}
                \subfloat{
                    \hspace{2.5cm}
                    \begin{tikzpicture}[scale=0.6, transform shape]
    \definecolor{beryllium}{RGB}{192,192,192} 
    \definecolor{hydrogen}{RGB}{173,216,230} 

    \draw[thick] (-2.7,0) -- (0,0);
    \draw[thick] (0,0) -- (2.7,0);
    
    \shade[ball color=hydrogen] (-2.7,0) circle (0.5);
    \node at (-2.7,0) {\textbf{H}};
    
    \shade[ball color=beryllium] (0,0) circle (0.7);
    \node at (0,0) {\textbf{Be}};
    
    \shade[ball color=hydrogen] (2.7,0) circle (0.5);
    \node at (2.7,0) {\textbf{H}};

    \draw[<->] (0,-0.9) -- (2.7,-0.9) node[midway, below] {2.5065 a.u.};

\end{tikzpicture}
                    \vspace{0.35cm}
                }\\
                \subfloat{
                    \def\arraystretch{1.5}
                     \resizebox{\linewidth}{!}{%
                    \begin{tabular}{|c|c|c|c|c|}
                        \hline
                        Starting basis & HF & \ac{DMRG} & \ac{FCI} & MRA+\ac{DMRG} \\
                        \hline\hline
                        STO-6G (7 orbitals) & -15.7245750 & -15.7595891 & -15.7595891 & -15.8065347 \\
                        (15 orbitals) & -- & -- & -- & -15.8801641 \\
                        cc-pVDZ (24 orbitals) & -15.7672727 & -15.8357162 & -15.8363604 & -- \\
                        \hline
                    \end{tabular}}
                }
            \end{minipage}
            \vspace{1.26cm}
        }
    }\\
\end{figure*}
\begin{figure*}
    \ContinuedFloat
    \subfloat{%
        \label{fig:N2_results}
        \subfloat{%
            \includegraphics[width=0.48\linewidth]{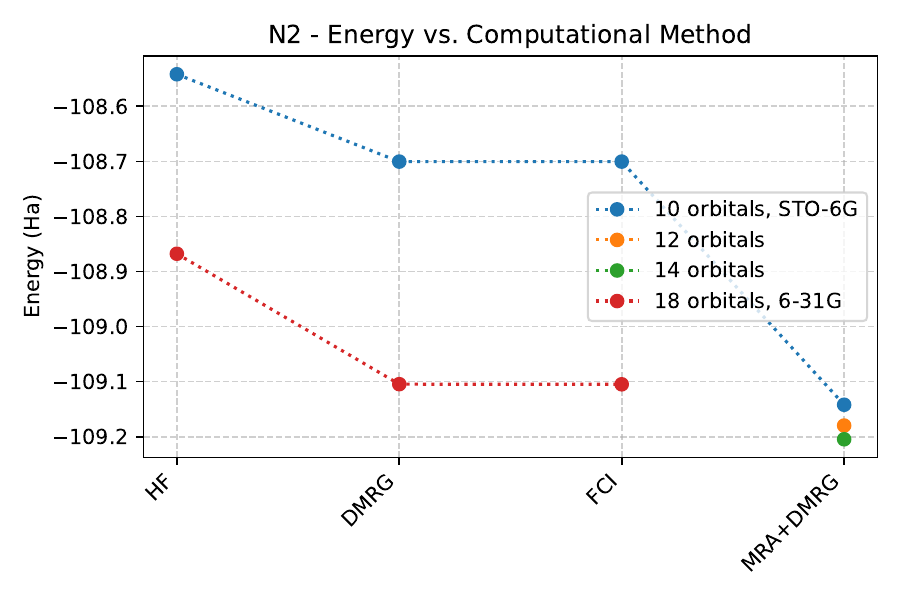}%
        }
        \subfloat{%
             \begin{minipage}{0.5\linewidth}
                \subfloat{
                    \hspace{3.2cm}
                    \begin{tikzpicture}[scale=0.6, transform shape]
    \definecolor{nitrogen}{RGB}{30,144,255} 

    \draw[thick] (0.4,0.2) -- (2,0.2);
    \draw[thick] (0.4,0) -- (2,0);
    \draw[thick] (0.4,-0.2) -- (2,-0.2);
    
    \shade[ball color=nitrogen] (0,0) circle (0.8);
    \node at (0,0) {\textbf{N}};
    
    \shade[ball color=nitrogen] (2.2,0) circle (0.8);
    \node at (2.2,0) {\textbf{N}};
    
    \draw[<->] (0,-1) -- (2.2,-1) node[midway, below] {2.074 a.u.};
\end{tikzpicture}
                    \vspace{0.1cm}
                }\\
                \subfloat{
                    \def\arraystretch{1.5}
                     \resizebox{\linewidth}{!}{%
                    \begin{tabular}{|c|c|c|c|c|}
                        \hline
                        Starting basis & HF & \ac{DMRG} & \ac{FCI} & MRA+\ac{DMRG} \\
                        \hline\hline
                        STO-6G (10 orbitals) & -108.541775 & -108.700424 & -108.700424 & -109.142037 \\
                        (12 orbitals) & -- & -- & -- & -109.179847 \\
                        (14 orbitals) & -- & -- & -- &  -109.204704\\
                        6-31G (18 orbitals) & -108.867774 & -109.1046517 & -109.1048996 &  -- \\
                        \hline
                    \end{tabular}}
                }
            \end{minipage}
            \vspace{1.25cm}
        }
    }\\
    \caption{Final energies obtained from Hartree Fock, \ac{DMRG}, \ac{FCI}, and our method combining \ac{DMRG} and MRA. When using MRA, a precision of $10^{-5}$ has been considered with polynomial order $k=9$. The MRA+\ac{DMRG} method works in principle with any number of orbitals. However, the current pilot implementation is limited to 15 orbitals. For \ac{AO} calculations, the number of orbitals is dictated by the choice of basis. For these reasons, the tables present empty entries.}
    \label{fig:numerical_results}
\end{figure*}

The results can be found in Fig.~\ref{fig:numerical_results} and among all the molecules we have tested, we can provide the following remarks:
\begin{itemize}
    \item MRA eliminates the need for a predefined atomic basis. This allows for calculations with any specified number of orbitals, for which there might be no corresponding atomic orbital basis set.
    \item The simple \ac{DMRG} calculation matches with \ac{FCI}, often with precision beyond the 8th decimal digit. For such small systems, this was entirely expected and serves as a benchmark for \texttt{chemtensor}.
    \item In every scenario, and while maintaining a fixed number of orbitals, our algorithm shows significant improvements in the final energy. This effect is especially evident in the \ce{N2} example with 10 orbitals, where the MRA+\ac{DMRG} algorithm yields a $442\text{ mHa}$ energy reduction compared to \ac{FCI} using a fixed atomic orbital basis set.
    \item Sometimes, the MRA+\ac{DMRG} algorithm outperforms the standard \ac{FCI} on an atomic orbital basis set even when increasing the basis size of the latter.
    Considering again the \ce{N2} example, the 10 and 12 orbitals' final results are still more accurate than \ac{FCI} with $6$-$31$G basis (and 18 orbitals).
    Further evidence of this effect can be seen in the \ce{BeH2} calculations, where the 15 orbitals' MRA+\ac{DMRG} final energy constitutes a $44\text{ mHa}$ improvement compared to the \ac{FCI} calculation with the cc-pVDZ basis (and 24 orbitals).
    \item As expected, such improvements tend to saturate when increasing the number of orbitals. This can be seen, for example, in the \ce{H2} simulation, where the energies of MRA+\ac{DMRG} with 10 and 12 orbitals and \ac{FCI} with 28 orbitals differ by less than $1\text{ mHa}$.
\end{itemize}
These results reflect the fact that our method works at the \ac{CBS} limit, up to any predefined precision $p$, while the standard implementation of \ac{HF}, \ac{DMRG} and \ac{FCI} relies on a finite atomic orbital basis set.
Notably, the computational scaling of \ac{DMRG} depends only on the number of orbitals, not on the basis size.
As a result, our method holds the promise of improving \ac{DMRG} by increasing the basis size up to the \ac{CBS} limit while preserving the number of orbitals and, thus, limiting its computational complexity.

In Appendix~\ref{appendix_plots} we have reported some of the convergence plots for the molecules tested in this section. For such systems, we observed smooth convergence and the energy parameters defining the Helmholtz kernel in Eq.~\eqref{eq_sign_Helmholtz_coefficients} are found to be negative. This aligns with the empirical observation that $\varepsilon^\prime_{mm}/\Lambda_m$ tends to be negative, a necessary condition for the algorithm to function correctly.

It is worth noting that the \ac{MRA}+\ac{DMRG} algorithm qualifies as an ab initio method. While it is theoretically possible to combine it with methods such as Hartree-Fock or DFT, we did not need to perform any preliminary approximate calculation in the considered systems.
Moreover, our method seems to have only a weak dependence on the choice of initial orbitals.
We have used different Gaussian basis sets to build the starting wavefunctions, as well as Slater-type orbitals, but no practical difference has been observed in the final energies and the convergence paths.
While this remains to be tested on larger systems, it appears to be a promising feature of such an ab initio method.

\begin{figure}[t]
    \centering
    ~\hfill
    \subfloat[H2 dissociation]{%
        \label{fig:H2_convergence}
        \includegraphics[width=0.48\linewidth]{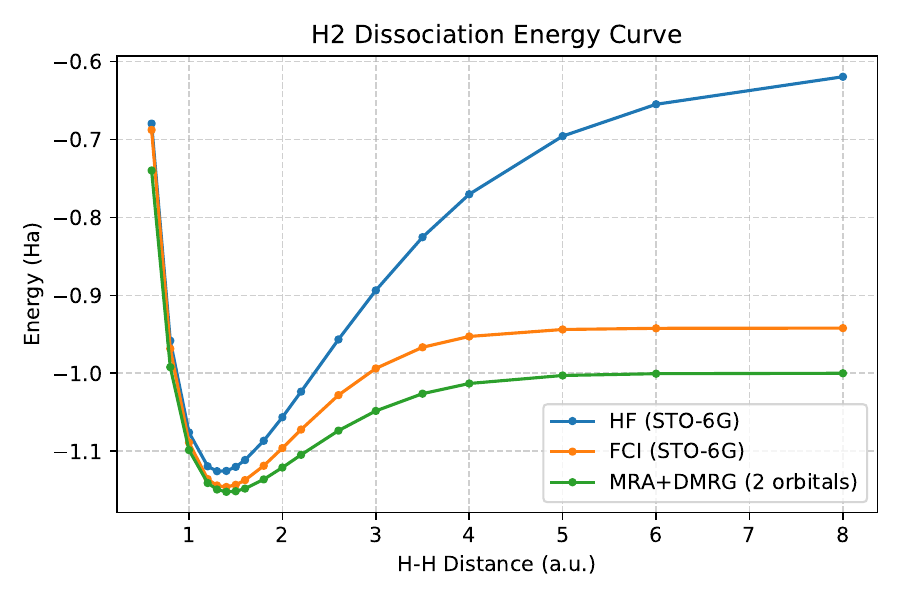}%
    }\hfill
    \subfloat[N2 dissociation]{%
        \label{fig:N2_convergence}
        \includegraphics[width=0.48\linewidth]{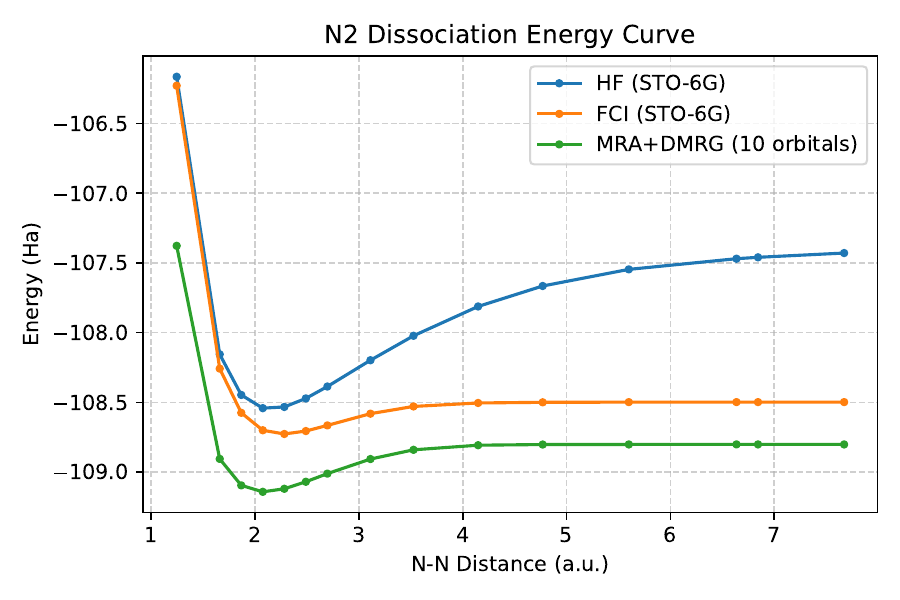}%
    }\hfill~
    \caption{Dissociation paths of H2 and N2 with ground state energies obtained via HF and \ac{FCI} with a minimal basis set and the MRA+\ac{DMRG} algorithm with the same number of orbitals, respectively.}
    \label{fig:H2_dissociation}
\end{figure}

Additionally, in Fig.~\ref{fig:H2_dissociation}, we have plotted the Hydrogen and Nitrogen dissociation energy paths as a further benchmark of our method: they show the correct asymptotic behavior that one expects from the proper multiconfigurational treatment. 

\section{Conclusion}
This work represents a first attempt to apply \ac{DMRG} within the MRA and \acp{MW} framework.
Specifically, we have developed an ab initio and self-consistent orbital optimization algorithm inspired by the work of Valeev et al.~\cite{Valeev_2023}. 
While following a similar workflow, our method replaces CI energy estimates with \ac{DMRG} and extracts energy gradients directly from the \ac{DMRG} tensor network, avoiding the need for \ac{RDM} evaluations, as described in Appendix~\ref{appendix_DMRG}.

We have tested our approach on small molecules with up to 15 orbitals, with results presented in Section~\ref{sec_numerical_results}. 
Our method consistently achieves lower energies for a fixed number of orbitals than standalone \ac{FCI} or \ac{DMRG}, benefiting from the larger basis set. 
This demonstrates that the combination of MRA and \ac{DMRG} successfully captures the multireference nature of the system while approaching the infinite basis set limit, with MRA precision $p$ and the number of orbitals as the primary limitations.

Despite its potential, our approach is currently limited to small systems with a restricted number of orbitals due to computational limitations.
For the MRA calculations, we relied on the Python package \vampyr{} \cite{vampyr_2024}, which, while user-friendly, is not optimized for an \ac{HPC} framework.
Future work will focus on migrating to the more efficient \mrchem{}~\cite{mrchem_2023} and optimizing \texttt{chemtensor}~\cite{chemtensor} for \ac{HPC} integration.  

As mentioned in Appendix~\ref{appendix_plots}, all cases considered show a smooth convergence up to energy precision $\delta=10^{-5} \text{ Ha}$.
In the future, we aim to consider larger systems, which may exhibit slower or more difficult convergence.
In this regard, the algorithm could be improved through a Newton optimization method~\cite{dinvay2025newton}, which would further reduce its dependence on the initial guess.

Another unexplored aspect is orbital localization.
Localized orbitals yield a more compact MRA representation, reducing the computational cost. \ac{DMRG} calculations also benefit from localization in strongly correlated systems \cite{Olivares-Amaya_2015}. For these reasons, incorporating a localization step into the algorithm could enhance its overall efficiency.
At present, however, it is unclear to what extent localization can be exploited. The current algorithm is designed around the diagonalization of the one-body \ac{RDM}, which restricts the orbital basis to the \acp{NO} \cite{loewdin1954quantum-03a}. This constraint is absent in standard \ac{SCF} calculations, where localization can be applied more freely. \Aclp{NO} are not necessarily localized, although adaptations to improve locality have been proposed \cite{reed1985natural-eb3}.
Moreover, recent work by one of the authors \cite{dinvay2025heat} has shown that the explicit application of the Laplacian operator, which is currently avoided by diagonalizing the one-body \ac{RDM}, can be performed efficiently. This would allow different orbital choices, thus enabling a more systematic exploitation of locality.

During this project, we became aware of the work by Langkabel et al.~\cite{mw_vqe}, who also built upon the Lagrangian optimization from  Valeev et al.~\cite{Valeev_2023} but replaced CI calculations with a variational quantum eigensolver (VQE) \cite{Peruzzo_2014}.  
In future work, it would be interesting to compare these two techniques and explore their respective regimes of applicability.
This also highlights the growing interest in MRA and \acp{MW} techniques within different branches of quantum chemistry.

Our work is an initial step in bridging two distinct approaches: multiresolution analysis and tensor network applications in chemistry. 
The presented results seem promising; however, further optimization is needed to assess the algorithm's feasibility in larger and more complex systems.

\section*{Acknowledgements}
LF would like to thank Simen Kvaal for enabling this work by putting us in contact with CM. Dr. Roberto Di Remigio Eikås is also acknowledged for valuable discussions.

MN acknowledges funding by the Munich Quantum Valley, section K5 Q-DESSI. The research is part of the Munich Quantum Valley, which is supported by the Bavarian state government with funds from the Hightech Agenda Bayern Plus.

The financial support from the Research Council of Norway through its Centres of Excellence scheme (Hylleraas centre, 262695) is acknowledged. The support from the Norwegian Metacenter for Computational Science (NOTUR) infrastructure under the nn14654k grant of computer time is also acknowledged.

\appendix
\section{DMRG and gradients}
\label{appendix_DMRG}
The \ac{DMRG} algorithm \cite{White_1992, Schollwoeck_2011} is a tensor network method suitable for approximating the ground state of (strongly correlated) quantum systems in condensed matter physics and chemistry \cite{White_1999, Chan_2002, Chan_2011, Olivares-Amaya_2015, Chan_2016, Ma_2022}. \ac{DMRG} uses a \ac{MPS} as a variational Ansatz for the wavefunction and can reach energies close to the exact (\ac{FCI}) ground state.
For a fixed virtual bond dimension, DMRG exhibits a computational cost that scales polynomially with the number of orbitals. While the required bond dimension remains modest in many practical scenarios, this is not guaranteed, and its growth in strongly correlated systems is still poorly understood and difficult to predict.

The molecular Hamiltonian in Eq.~\eqref{eq_def_second_quantization_Hamiltonian} exhibits abelian symmetries corresponding to global particle number and spin conservation. In general, such abelian symmetries endow the \ac{MPS} and \ac{MPO} tensors with a block sparsity structure, which can be advantageous to increase computational performance \cite{Singh_2011}. In the current setting, we associate a particle number and spin tuple $(q_n, q_s)$ with each index at every tensor leg. The tensor legs also have a predefined direction, either outgoing or incoming. The conservation laws then require that the sum of incoming quantum numbers must be equal to the sum of outgoing quantum numbers; otherwise, the respective tensor entry has to be zero. We use the convention that each local ``site'' $\ell$ has dimension $4$, corresponding to the possible occupation patterns of the two spin orbitals $\{\phi_{\ell, \uparrow}, \phi_{\ell, \downarrow}\}$, and enumerate these patterns as follows:
\begin{center}
\setlength\tabcolsep{2mm}
\renewcommand*{\arraystretch}{1.5}
\begin{tabular}{c|cccc}
index & 0 & 1 & 2 & 3\\
\hline
$\{\phi_{\ell, \uparrow}, \phi_{\ell, \downarrow}\}$ & \emptyorb\,\emptyorb & \emptyorb\,\fullorb & \fullorb\,\emptyorb & \fullorb\,\fullorb\\
$q_n$ & $0$ & $1$ & $1$ & $2$\\
$q_s$ & $0$ & $-\frac{1}{2}$ & $\frac{1}{2}$ & $0$
\end{tabular}
\end{center}
The sparsity condition of the tensors can be illustrated for an individual \ac{MPS} tensor:
\[
\begin{tikzpicture}[style={inner sep=0}, >=stealth, decoration={markings,mark=at position 0.5 with {\arrow{>}}}]
\node[draw,circle,thick,minimum size=28] (a) at (0, 0) {$A$};
\node at (-0.3, 0) {\color{blue}\scriptsize $0$};
\node at ( 0, 0.3) {\color{blue}\scriptsize $1$};
\node at ( 0.3, 0) {\color{blue}\scriptsize $2$};
\draw[postaction={decorate}] (a) --node[above=1mm] {\footnotesize $i$} (-1, 0);
\draw[postaction={decorate}] (a) --node[right=1mm] {\footnotesize $j$} ( 0, 1);
\draw[postaction={decorate}] ( 1, 0) --node[above=1mm] {\footnotesize $k$} (a);
\end{tikzpicture}
\]
The inner blue numbers define the tensor leg order. The horizontal legs ($0$ and $2$) are the virtual bonds, and the vertical leg ($1$) is the physical dimension. Each leg stores a separate list of quantum number tuples. For example, for the physical leg, these are $\{(q_{n,j}^{(1)}, q_{s,j}^{(1)})\}_{j=0,\dots,3} = \{ (0, 0), (1, -\frac{1}{2}), (1, \frac{1}{2}), (2, 0)\}$ according to the above table, where the superscript indicates the tensor leg. The sparsity condition means that the entry $A_{ijk}$ can be non-zero only if
\begin{subequations}
\begin{align}
q_{n,i}^{(0)} + q_{n,j}^{(1)} &= q_{n,k}^{(2)},\\
q_{s,i}^{(0)} + q_{s,j}^{(1)} &= q_{s,k}^{(2)},
\end{align}
\end{subequations}
i.e., if the quantum number ``inflow'' matches the ``outflow''.

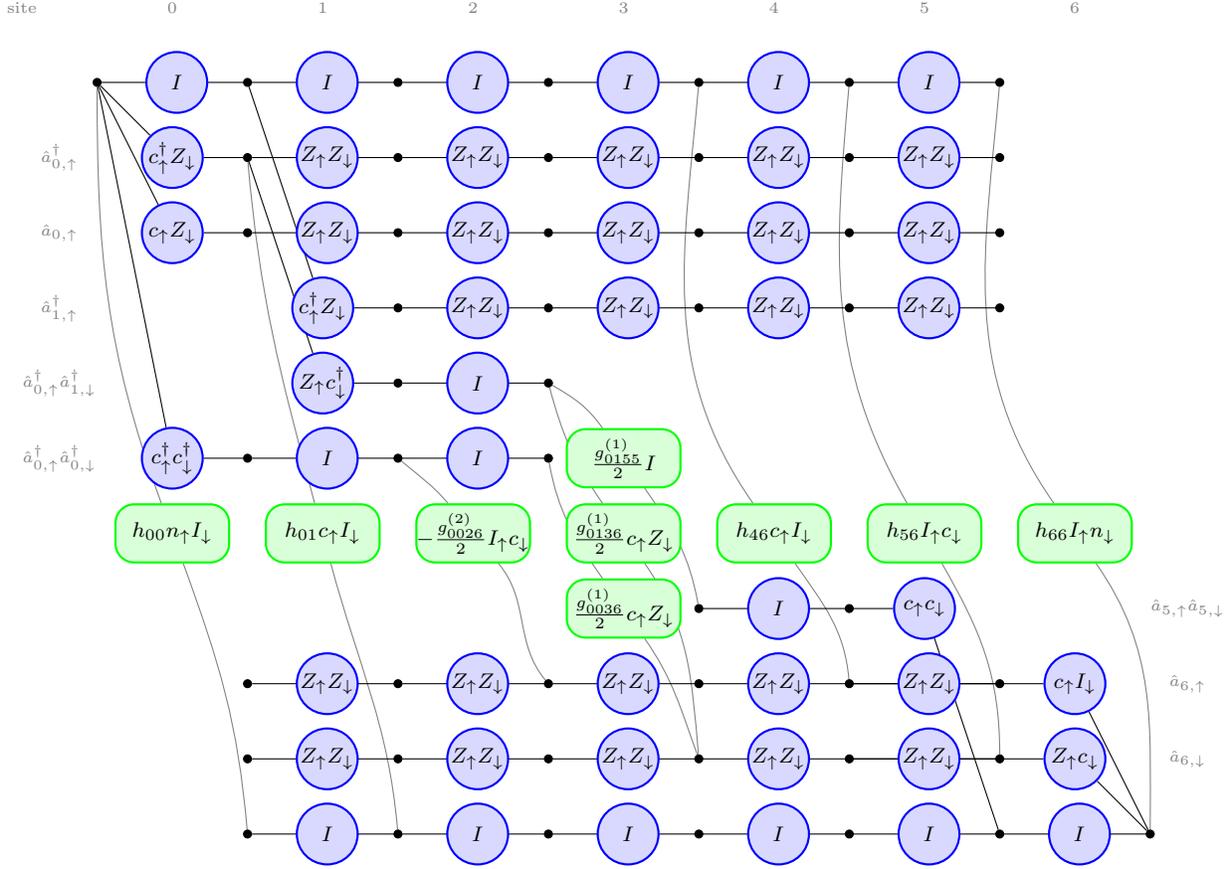
\begin{figure*}[t]
\centering
\begin{tikzpicture}[style={inner sep=0}]
\tikzset{vertex style/.style={draw, circle, fill=black, minimum size=3}}
\tikzset{scaffold op style/.style={draw=blue, thick, circle, fill=blue!15!white, minimum size=23}}
\tikzset{term op style/.style={draw=green, thick, rectangle, rounded corners=.25cm, fill=green!15!white, minimum height=22, minimum width=43}}

\node[gray] at (-1, 11) {\scriptsize site};
\foreach \i in {0,...,6} {
    \node[gray] at (2*\i+1, 11) {\scriptsize $\i$};
}
\foreach \i in {0,...,6} {
    \node[vertex style] (vidl\i) at (2*\i, 10) {};
}
\foreach \i in {0,...,5} {
    \pgfmathsetmacro{\ip}{\i+1}
    \draw (vidl\i) --node[scaffold op style, label=center:\scriptsize $I$] {} (vidl\ip);
}
\foreach \i in {1,...,7} {
    \node[vertex style] (vidr\i) at (2*\i, 0) {};
}
\foreach \i in {1,...,6} {
    \pgfmathsetmacro{\ip}{\i+1}
    \draw (vidr\i) --node[scaffold op style, label=center:\scriptsize $I$] {} (vidr\ip);
}
\node[gray] at (-0.5, 9) {\scriptsize $\hat{a}^{\dagger}_{0,\uparrow}$};
\foreach \i in {1,...,6} {
    \node[vertex style] (vad0up\i) at (2*\i, 9) {};
}
\node[scaffold op style, label=center:\scriptsize $c^{\dagger}_{\uparrow} Z_{\downarrow}$] (ead0up) at (1, 9) {};
\draw (vidl0) -- (ead0up) -- (vad0up1);
\foreach \i in {1,...,5} {
    \pgfmathsetmacro{\ip}{\i+1}
    \draw (vad0up\i) --node[scaffold op style, label=center:\scriptsize $Z_{\uparrow} Z_{\downarrow}$] {} (vad0up\ip);
}
\node[gray] at (-0.5, 8) {\scriptsize $\hat{a}_{0,\uparrow}$};
\foreach \i in {1,...,6} {
    \node[vertex style] (va0up\i) at (2*\i, 8) {};
}
\node[scaffold op style, label=center:\scriptsize $c_{\uparrow} Z_{\downarrow}$] (ea0up) at (1, 8) {};
\begin{scope}[on background layer]
\draw (vidl0) -- (ea0up) -- (va0up1);
\end{scope}
\foreach \i in {1,...,5} {
    \pgfmathsetmacro{\ip}{\i+1}
    \draw (va0up\i) --node[scaffold op style, label=center:\scriptsize $Z_{\uparrow} Z_{\downarrow}$] {} (va0up\ip);
}
\node[gray] at (-0.5, 7) {\scriptsize $\hat{a}^{\dagger}_{1,\uparrow}$};
\foreach \i in {2,...,6} {
    \node[vertex style] (vad1up\i) at (2*\i, 7) {};
}
\node[scaffold op style, label=center:\scriptsize $c^{\dagger}_{\uparrow} Z_{\downarrow}$] (ead1up) at (3, 7) {};
\begin{scope}[on background layer]
\draw (vidl1) -- (ead1up) -- (vad1up2);
\end{scope}
\foreach \i in {2,...,5} {
    \pgfmathsetmacro{\ip}{\i+1}
    \draw (vad1up\i) --node[scaffold op style, label=center:\scriptsize $Z_{\uparrow} Z_{\downarrow}$] {} (vad1up\ip);
}
\node[gray] at (-0.5, 6) {\scriptsize $\hat{a}^{\dagger}_{0,\uparrow} \hat{a}^{\dagger}_{1,\downarrow}$};
\foreach \i in {2,...,3} {
    \node[vertex style] (vad0upad1up\i) at (2*\i, 6) {};
}
\node[scaffold op style, label=center:\scriptsize $Z_{\uparrow} c^{\dagger}_{\downarrow}$] (ead1dn) at (3, 6) {};
\begin{scope}[on background layer]
\draw (vad0up1) -- (ead1dn) -- (vad0upad1up2);
\end{scope}
\foreach \i in {2,...,2} {
    \pgfmathsetmacro{\ip}{\i+1}
    \draw (vad0upad1up\i) --node[scaffold op style, label=center:\scriptsize $I$] {} (vad0upad1up\ip);
}
\node[gray] at (-0.5, 5) {\scriptsize $\hat{a}^{\dagger}_{0,\uparrow} \hat{a}^{\dagger}_{0,\downarrow}$};
\foreach \i in {1,...,3} {
    \node[vertex style] (vad0upad0dn\i) at (2*\i, 5) {};
}
\node[scaffold op style, label=center:\scriptsize $c^{\dagger}_{\uparrow} c^{\dagger}_{\downarrow}$] (ead0upad0dn) at (1, 5) {};
\begin{scope}[on background layer]
\draw (vidl0) -- (ead0upad0dn) -- (vad0upad0dn1);
\end{scope}
\foreach \i in {1,...,2} {
    \pgfmathsetmacro{\ip}{\i+1}
    \draw (vad0upad0dn\i) --node[scaffold op style, label=center:\scriptsize $I$] {} (vad0upad0dn\ip);
}
\node[gray] at (14.5, 3) {\scriptsize $\hat{a}_{5,\uparrow} \hat{a}_{5,\downarrow}$};
\foreach \i in {4,...,5} {
    \node[vertex style] (va5upa5dn\i) at (2*\i, 3) {};
}
\node[scaffold op style, label=center:\scriptsize $c_{\uparrow} c_{\downarrow}$] (ea5upa5dn) at (11, 3) {};
\begin{scope}[on background layer]
\draw (va5upa5dn5) -- (ea5upa5dn) -- (vidr6);
\end{scope}
\foreach \i in {4,...,4} {
    \pgfmathsetmacro{\ip}{\i+1}
    \draw (va5upa5dn\i) --node[scaffold op style, label=center:\scriptsize $I$] {} (va5upa5dn\ip);
}
\node[gray] at (14.5, 2) {\scriptsize $\hat{a}_{6,\uparrow}$};
\foreach \i in {1,...,6} {
    \node[vertex style] (va6up\i) at (2*\i, 2) {};
}
\node[scaffold op style, label=center:\scriptsize $c_{\uparrow} I_{\downarrow}$] (ea6up) at (13, 2) {};
\begin{scope}[on background layer]
\draw (va6up5) -- (ea6up) -- (vidr7);
\end{scope}
\foreach \i in {1,...,5} {
    \pgfmathsetmacro{\ip}{\i+1}
    \draw (va6up\i) --node[scaffold op style, label=center:\scriptsize $Z_{\uparrow} Z_{\downarrow}$] {} (va6up\ip);
}
\node[gray] at (14.5, 1) {\scriptsize $\hat{a}_{6,\downarrow}$};
\foreach \i in {1,...,6} {
    \node[vertex style] (va6dn\i) at (2*\i, 1) {};
}
\node[scaffold op style, label=center:\scriptsize $Z_{\uparrow} c_{\downarrow}$] (ea6dn) at (13, 1) {};
\begin{scope}[on background layer]
\draw (va6dn5) -- (ea6dn) -- (vidr7);
\end{scope}
\foreach \i in {1,...,5} {
    \pgfmathsetmacro{\ip}{\i+1}
    \draw (va6dn\i) --node[scaffold op style, label=center:\scriptsize $Z_{\uparrow} Z_{\downarrow}$] {} (va6dn\ip);
}

\node[term op style, label=center:\scriptsize $h_{00} n_{\uparrow} I_{\downarrow}$] (en0up) at (1, 4) {};
\begin{scope}[on background layer]
\draw[gray] (vidl0) to[out=-90, in=110] (en0up) to[out=-70, in=95] (vidr1);
\end{scope}
\node[term op style, label=center:\scriptsize $h_{01} c_{\uparrow} I_{\downarrow}$] (ead1up) at (3, 4) {};
\begin{scope}[on background layer]
\draw[gray] (vad0up1) to[out=-85, in=105] (ead1up) to[out=-75, in=95] (vidr2);
\end{scope}
\node[term op style, label=center:\scriptsize $-\frac{g^{(2)}_{0026}}{2}  I_{\uparrow} c_{\downarrow}$] (ead2up) at (5, 4) {};
\begin{scope}[on background layer]
\draw[gray] (vad0upad0dn2) to[out=-40, in=130] (ead2up) to[out=-50, in=130] (va6up3);
\end{scope}
\node[term op style, label=center:\scriptsize $h_{46} c_{\uparrow} I_{\downarrow}$] (ead4up) at (9, 4) {};
\begin{scope}[on background layer]
\draw[gray] (vidl4) to[out=-95, in=125] (ead4up) to[out=-55, in=95] (va6up5);
\end{scope}
\node[term op style, label=center:\scriptsize $h_{56} I_{\uparrow} c_{\downarrow}$] (ead5dn) at (11, 4) {};
\begin{scope}[on background layer]
\draw[gray] (vidl5) to[out=-95, in=120] (ead5dn) to[out=-60, in=90] (va6dn6);
\end{scope}
\node[term op style, label=center:\scriptsize $h_{66} I_{\uparrow} n_{\downarrow}$] (en6dn) at (13, 4) {};
\begin{scope}[on background layer]
\draw[gray] (vidl6) to[out=-95, in=125] (en6dn) to[out=-55, in=90] (vidr7);
\end{scope}
\node[term op style, label=center:\scriptsize $\frac{g^{(1)}_{0155}}{2}  I$] (eid3) at (7, 5) {};
\begin{scope}[on background layer]
\draw[gray] (vad0upad1up3) to[out=-30, in=125] (eid3) to[out=-55, in=100] (va5upa5dn4);
\end{scope}
\node[term op style, label=center:\scriptsize $\frac{g^{(1)}_{0136}}{2}  c_{\uparrow} Z_{\downarrow}$] (ea3up) at (7, 4) {};
\begin{scope}[on background layer]
\draw[gray] (vad0upad1up3) to[out=-75, in=125] (ea3up) to[out=-55, in=95] (va6dn4);
\end{scope}
\node[term op style, label=center:\scriptsize $\frac{g^{(1)}_{0036}}{2}  c_{\uparrow} Z_{\downarrow}$] (ea3up) at (7, 3) {};
\begin{scope}[on background layer]
\draw[gray] (vad0upad0dn3) to[out=-80, in=125] (ea3up) to[out=-55, in=110] (va6dn4);
\end{scope}

\end{tikzpicture}
\caption{State diagram for constructing the \ac{MPO} representation of a molecular Hamiltonian \eqref{eq_def_second_quantization_Hamiltonian_mod}. $c^{\dagger}_{\sigma}$ and $c_{\sigma}$ for $\sigma \in \{ \uparrow, \downarrow \}$ are the local bosonic creation and annihilation operators, $Z_{\sigma}$ is the Pauli-$Z$ gate acting on spin sector $\sigma \in \{ \uparrow, \downarrow \}$, and $I$ is the identity operation. The ```scaffolding'' is highlighted in blue. The figure only shows a few operator strings and Hamiltonian terms for visual clarity.}
\label{fig:molecular_hamiltonian_mpo_construction}
\end{figure*}

Constructing an \ac{MPO} representation of the molecular Hamiltonian with optimal virtual bond dimension (scaling as $\mathcal{O}(L^2)$ for $L$ orbitals) is intricate in detail. A recipe reported in the literature is the so-called complementary operator approach \cite{Chan_2016}. Here, we sketch an alternative (but equivalent) viewpoint and procedure, illustrated in Fig.~\ref{fig:molecular_hamiltonian_mpo_construction}. The idea is to construct a ``scaffolding'' of creation and annihilation operator strings, then synthesize the Hamiltonian terms by connecting these strings with appropriate local operators. The graph-like data structure, denoted state diagram and described in \cite{Crosswhite_2008, Milbradt_2024_State}, helps organize this construction and track terms. The black dots in Fig.~\ref{fig:molecular_hamiltonian_mpo_construction} are interpreted as vertices of a graph; each vertex becomes a virtual bond in the final \ac{MPO}. The circles are the graph's edges and contain local operators, which populate the final \ac{MPO} tensors. We use the Jordan-Wigner transformation to account for the fermionic anticommutation relations. In practice, this means inserting Pauli-$Z$ gates between odd occurrences of the creation and annihilation operators.

More in detail, we can first combine the terms appearing in the interaction part of the Hamiltonian \eqref{eq_def_second_quantization_Hamiltonian} as follows:
\begin{equation}
\begin{split}
\label{eq_def_second_quantization_Hamiltonian_mod}
\hat{H} &= \sum_{i,j} \sum_{\sigma \in \{\uparrow, \downarrow\}} h_{ij} \, \hat{a}^{\dagger}_{i,\sigma} \hat{a}_{j,\sigma} \\
&+\frac{1}{2} \sum_{\substack{(i,\sigma) < (j,\tau),\\ (k,\sigma') < (\ell,\tau')}} \left(g^{(1)}_{ijk\ell} \, \delta_{\sigma,\sigma'} \delta_{\tau,\tau'} - g^{(2)}_{ijk\ell} \, \delta_{\sigma,\tau'} \delta_{\tau,\sigma'}  \right) \hat{a}^{\dagger}_{i,\sigma} \hat{a}^{\dagger}_{j,\tau} \hat{a}_{\ell,\tau'} \hat{a}_{k,\sigma'},
\end{split}
\end{equation}
where $(i,\sigma) < (j,\tau)$ is understood as lexicographical ordering, and the new coefficient tensors are:
\begin{subequations}
\begin{align}
    g^{(1)}_{ijk\ell} &= g_{i j k \ell} + g_{j i \ell k},\\
    g^{(2)}_{ijk\ell} &= g_{j i k \ell} + g_{i j \ell k}.
\end{align}
\end{subequations}
We remark that the representation in Eq.~\eqref{eq_def_second_quantization_Hamiltonian_mod} only uses the anti-commuting property of the fermionic operators and does not require any symmetry assumptions on the $g_{i j k \ell}$ tensor.

The state diagram in Fig.~\ref{fig:molecular_hamiltonian_mpo_construction} contains two terminal vertices on the left and right end, corresponding to (dummy) virtual bonds of dimension $1$. The scaffolding consists of:
\begin{itemize}
    \item Identity strings that start from the terminal vertices on either end and run to the other side.
    \item Single creation and annihilation operators $\hat{a}^{\dagger}_{i,\sigma}$ and $\hat{a}_{i,\sigma}$ for all $i \in \{ 0, \dots, L - 1 \}$ and $\sigma \in \{ \uparrow, \downarrow \}$, starting from both sides. To facilitate connections to other operators, we extend the strings throughout the system (using local $Z$ operators based on the Jordan-Wigner transform). For example, $\hat{a}^{\dagger}_{1,\uparrow}$ starting from the left is represented as:
    \begin{equation}
        \hat{a}^{\dagger}_{1,\uparrow} = I \otimes c^{\dagger}_{\uparrow} Z_{\downarrow} \otimes Z_{\uparrow} Z_{\downarrow} \otimes Z_{\uparrow} Z_{\downarrow} \otimes \cdots,
    \end{equation}
    where $c^{\dagger}_{\uparrow}$ is the local (bosonic) creation operator acting on the spin-up sector. Explicitly,
    \begin{equation}
        c_{\sigma} = \begin{pmatrix} 0 & 1 \\ 0 & 0 \end{pmatrix}, \quad%
        c^{\dagger}_{\sigma} = \begin{pmatrix} 0 & 0 \\ 1 & 0 \end{pmatrix}, \quad%
        \sigma \in \{ \uparrow, \downarrow \}.
    \end{equation}
    The enumeration of occupation patterns at site $\ell$ described above implies that products of local operators acting on the spin-up and spin-down sectors, like $c^{\dagger}_{\uparrow} Z_{\downarrow}$, are Kronecker products: $c^{\dagger}_{\uparrow} Z_{\downarrow} = \left(\begin{smallmatrix} 0 & 0 \\ 1 & 0 \end{smallmatrix}\right) \otimes \left(\begin{smallmatrix} 1 & 0 \\ 0 & -1 \end{smallmatrix}\right)$. The operator string of $\hat{a}^{\dagger}_{i,\sigma}$ and $\hat{a}_{i,\sigma}$ is connected to the identity string of its originating direction, thus avoiding a separate copy of local identity operators.
    \item Quadratic fermionic operators: $\hat{a}^{\dagger}_{i,\sigma} \hat{a}^{\dagger}_{j,\tau}$ for $(i, \sigma) < (j, \tau)$, $\hat{a}_{j,\tau} \hat{a}_{i,\sigma}$ for $(i, \sigma) < (j, \tau)$, and $\hat{a}^{\dagger}_{i,\sigma} \hat{a}_{j,\tau}$. For all of these, the orbital indices $i$ and $j$ are both either in the left half $\{ 0, \dots, \floor{\frac{L}{2}} - 1 \}$ or both in the right half $\{ \floor{\frac{L}{2}} + 1, \dots L - 1 \}$, where $\floor{\cdot}$ is the floor function. Fig.~\ref{fig:molecular_hamiltonian_mpo_construction} shows some instances of the quadratic fermionic operator strings.
\end{itemize}
The scaffolding contains all state diagram vertices required for the \ac{MPO} construction and thus also determines the virtual bond dimensions.

In the second phase, one integrates the Hamiltonian terms by adding and connecting local operators, shown in green in Fig.~\ref{fig:molecular_hamiltonian_mpo_construction}. For example, the kinetic term $h_{01} \hat{a}^{\dagger}_{0,\uparrow} \hat{a}_{1,\uparrow}$ is implemented by adding a state diagram edge containing the local operator $h_{01} c_{\uparrow} I_{\downarrow}$ at site $1$ and connecting this edge to the closest vertex of the string for $\hat{a}^{\dagger}_{0,\uparrow}$ on the left and to the closest vertex of the identity string attached to the right terminal on the right. Note that the Hamiltonian coefficients are solely contained in the (green) local operators added during the second phase.

During the two-site \ac{DMRG} sweep for optimizing a neighboring pair of \ac{MPS} tensors, the computational cost for applying the local effective Hamiltonian scales as $\mathcal{O}(M^2 D^2 + M^3 D)$ \emph{when assuming dense tensors}, where $M$ and $D$ denote the maximum bond dimensions of the \ac{MPS} and \ac{MPO}, respectively. For a molecular Hamiltonian and $L$ orbitals, $D = \mathcal{O}(L^2)$. A \ac{DMRG} sweep over all orbitals incurs another complexity factor of $L$, leading to an ostensible complexity scaling of $\mathcal{O}(M^2 L^5 + M^3 L^3)$. However, as noted in Ref.~\cite{Chan_2016}, the expensive $\mathcal{O}(M^2 L^5)$ computation is actually reduced to $\mathcal{O}(M^2 L^4)$ when exploiting the sparsity of the molecular \ac{MPO} tensors. In our current implementation \cite{chemtensor}, we exploit sparsity due to particle and spin conservation, but not yet the particular sparsity pattern due to a molecular Hamiltonian. We plan to add this feature in a future update of the code, and can use the existing data structures for the above state diagram construction for this purpose.

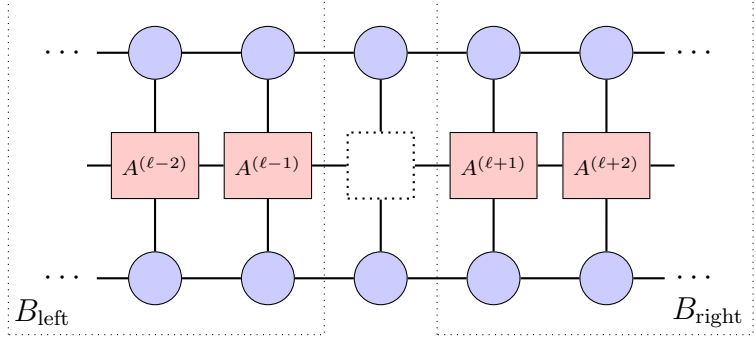
\begin{figure}[t]
    \centering
    \begin{tikzpicture}[scale=1.5]
    \tikzstyle{tensor} = [circle, draw, fill=blue!20, minimum size=20pt]
    \tikzstyle{operator} = [rectangle, draw, fill=red!20, minimum size=25pt, rounded corners=0pt]
    \tikzstyle{dotted_operator} = [rectangle, draw=black, thick, minimum size=25pt, dotted, rounded corners=0pt]

    \node[] (A0) at (-0.8,0) {$\cdots$};
    \node[tensor] (A1) at (0,0) {};
    \node[tensor] (A2) at (1,0) {};
    \node[tensor] (A3) at (2,0) {};
    \node[tensor] (A4) at (3,0) {};
    \node[tensor] (A5) at (4,0) {};
    \node[] (A6) at (4.8,0) {$\cdots$};

    \node[] (A0') at (-0.8,2) {$\cdots$};
    \node[tensor] (A1') at (0,2) {};
    \node[tensor] (A2') at (1,2) {};
    \node[tensor] (A3') at (2,2) {};
    \node[tensor] (A4') at (3,2) {};
    \node[tensor] (A5') at (4,2) {};
    \node[] (A6') at (4.8,2) {$\cdots$};
    
    \node[] (W0) at (-0.7,1) {};
    \node[operator] (W1) at (0,1) {\scriptsize $A^{(\ell-2)}$};
    \node[operator] (W2) at (1,1) {\scriptsize $A^{(\ell-1)}$};
    \node[dotted_operator] (W3) at (2,1) {};
    \node[operator] (W4) at (3,1) {\scriptsize $A^{(\ell+1)}$};
    \node[operator] (W5) at (4,1) {\scriptsize $A^{(\ell+2)}$};
    \node[] (W6) at (4.7,1) {};

    \draw[thick] (A0) -- (A1) -- (A2) -- (A3) -- (A4) -- (A5) -- (A6);
    \draw[thick] (A0') -- (A1') -- (A2') -- (A3') -- (A4') -- (A5') -- (A6');
    \draw[thick] (W0) -- (W1) -- (W2) -- (W3) -- (W4) -- (W5) -- (W6);
    \draw[thick] (A1') -- (W1) -- (A1);
    \draw[thick] (A2') -- (W2) -- (A2);
    \draw[thick] (A3') -- (W3) -- (A3);
    \draw[thick] (A4') -- (W4) -- (A4);
    \draw[thick] (A5') -- (W5) -- (A5);

    \draw[dotted] (-1.3, -0.5) rectangle (1.5, 2.5);
    \node at (-1, -0.3) {$B_{\text{left}}$};
    \draw[dotted] (2.5, -0.5) rectangle (5.3, 2.5);
    \node at (4.9, -0.3) {$B_{\text{right}}$};

\end{tikzpicture}
    \caption{Extraction of $\bra{\psi} \frac{\partial \hat{H}}{\partial A^{(\ell)}_J}\ket{\psi}$ for $\forall J$ for the energy gradients calculation. The environment blocks $B_{\text{left}}$ and $B_{\text{right}}$ can be precomputed for efficiency.}
    \label{fig:gradient}
\end{figure}

To compute energy gradients with respect to the Hamiltonian parameters, we first use the Hellmann-Feynman theorem: assuming that the state $\ket{\psi}$ represented as \ac{MPS} is an eigenstate of the Hamiltonian $\hat{H}$, then
\begin{subequations}
\begin{align}
    \frac{\partial E}{\partial h_{ij}} &= 
    \ev**{\frac{\partial \hat{H}}{\partial h_{ij}}}{\psi},\\
    \frac{\partial E}{\partial g_{ijk\ell}} &= \sum_{i',j',k',\ell'} 
    \ev**{\frac{\partial \hat{H}}{\partial g^{(1)}_{i'j'k'\ell'}}}{\psi} 
    \frac{\partial g^{(1)}_{i'j'k'\ell'}}{\partial g_{ijk\ell}} + 
    \ev**{\frac{\partial \hat{H}}{\partial g^{(2)}_{i'j'k'\ell'}}}{\psi} 
    \frac{\partial g^{(2)}_{i'j'k'\ell'}}{\partial g_{ijk\ell}}.
\end{align}
\end{subequations}
We evaluate the right sides using the tensor network representation. Denoting the $\ell$-th \ac{MPO} tensor of $\hat{H}$ by $A^{(\ell)}$ with entries addressed by a collective index $J$ (physical and virtual bond indices), then
\begin{equation}
    \ev**{\frac{\partial \hat{H}}{\partial h_{ij}}}{\psi} = \sum_{\ell=0}^{L-1} \sum_J \ev**{\frac{\partial \hat{H}}{\partial A^{(\ell)}_J}}{\psi} \frac{\partial A^{(\ell)}_J}{\partial h_{ij}}
\end{equation}
(and analogously for the interaction coefficients). The expression $\ev{\frac{\partial \hat{H}}{\partial A^{(\ell)}_J}}{\psi}$ (for all index values $J$) corresponds to omitting $A^{(\ell)}$ from the tensor network diagram representing $\bra{\psi} \hat{H} \ket{\psi}$, as schematically shown in Fig.~\ref{fig:gradient}. This gradient tensor can be efficiently computed, analogous to the effective Hamiltonian in DMRG. The derivative $ \frac{\partial A^{(\ell)}_J}{\partial h_{ij}}$ corresponds to one of the green local operators in Fig.~\ref{fig:molecular_hamiltonian_mpo_construction} without the coefficient $h_{ij}$. We have combined these steps for computing energy gradients in \texttt{chemtensor}\cite{chemtensor}.

The steps outlined here naturally follow from the chain rule and using the \ac{MPO} form of the Hamiltonian. It is well known that the energy gradients with respect to the single- and two-body coefficients are equal to the reduced density matrices, as described above in Eqs.~\eqref{eq:energy_gradients_rdms}. An algorithm for computing \acp{RDM} via tensor network methods has been developed in the literature before \cite{ZgidNooijen2008}, with computational complexity $\mathcal{O}(M^3 L^2 + M^2 L^4)$. Interestingly, one rediscovers their algorithm from our chain rule derivation, when additionally taking the state diagram construction in Fig.~\ref{fig:molecular_hamiltonian_mpo_construction} for the \ac{MPO} tensors $A^{(\ell)}$ into account. In particular, $\ev{\frac{\partial \hat{H}}{\partial A^{(\ell)}_J}}{\psi}$ does not need to be evaluated as a dense tensor; instead, only the entries corresponding to the green local operators are relevant, since they depend on the Hamiltonian coefficients.

\section{MRA+DMRG convergence plots} 
\label{appendix_plots}

\vspace{-5mm}
\begin{figure*}
    \centering
    \subfloat[Starting guess: Slater-type orbitals.]{%
        \label{fig:H2_convergence_SI}
        \includegraphics[width=0.46\linewidth]{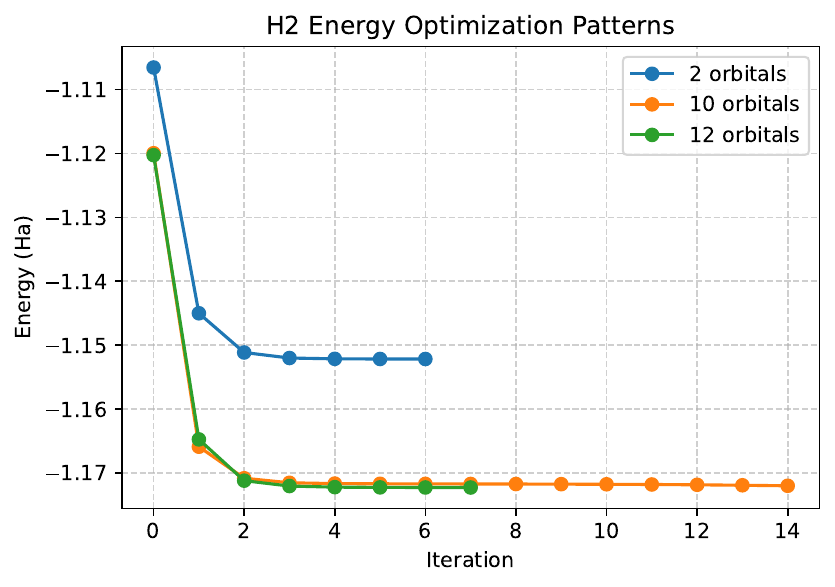}%
    }
    \subfloat[Starting guess: Slater-type orbitals.]{%
        \label{fig:He_convergence}
        \includegraphics[width=0.46\linewidth]{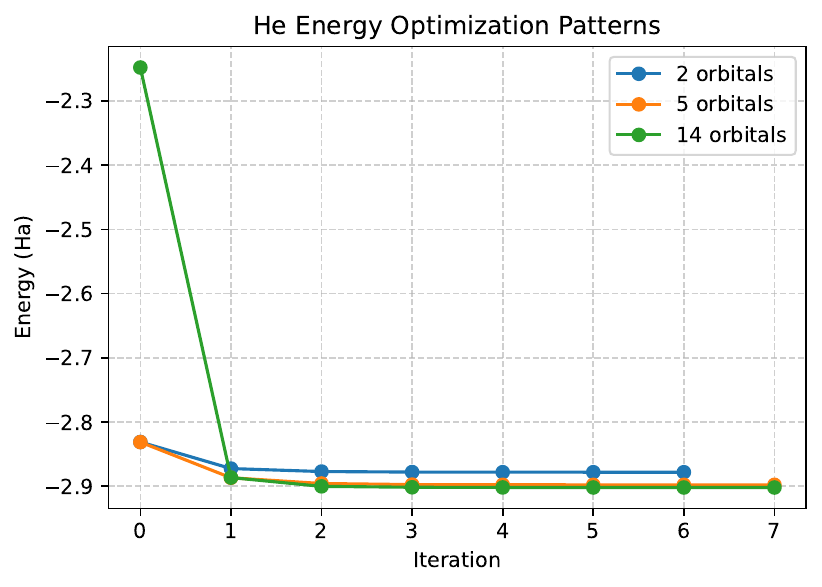}%
    }\\
    \subfloat[Starting guess: Slater-type orbitals for $9$ and $15$ orbitals. STO-$3$G for $3$ orbitals. Reduced cc-pVDZ (only $1$s and $2$s orbitals) for $6$ orbitals.]{%
        \label{fig:HeH2_convergence}
        \includegraphics[width=0.46\linewidth]{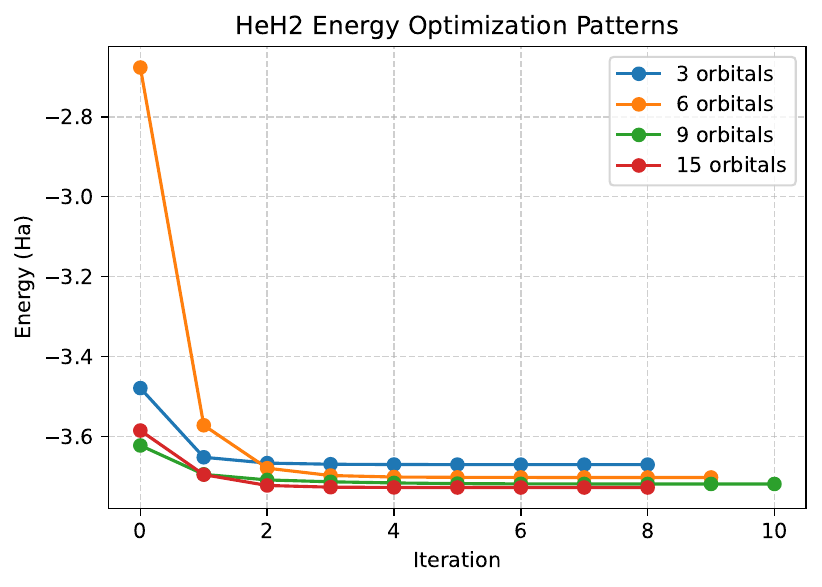}%
    }
    \subfloat[Starting guess: Slater-type orbitals.]{%
        \label{fig:BeH2_convergence}
        \includegraphics[width=0.46\linewidth]{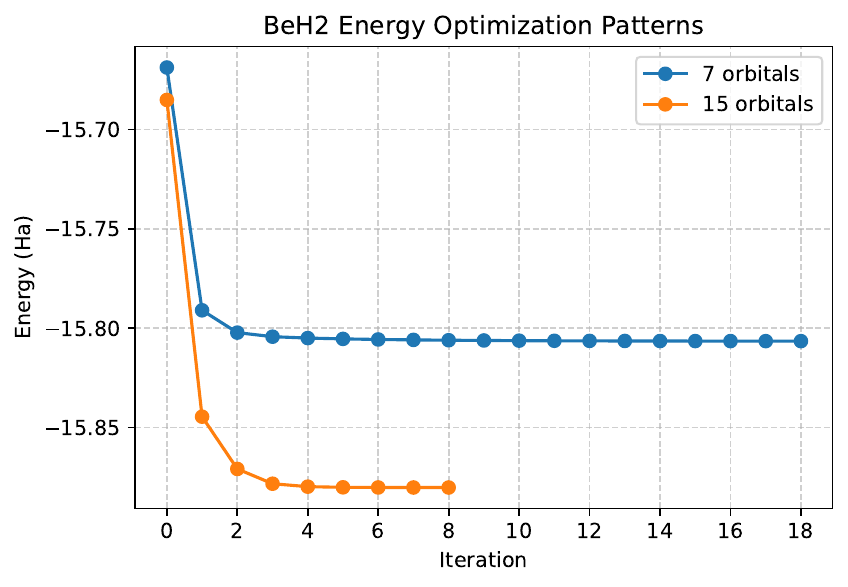}%
    }\\
    \subfloat[Starting guess: STO-$3$G for $10$ orbitals. Augmented STO-$3$G for $12$ orbitals. Augmented basis built from the previously converged $12$ orbitals for $14$ orbitals.]{%
        \label{fig:N2_convergence_SI}
        \includegraphics[width=0.46\linewidth]{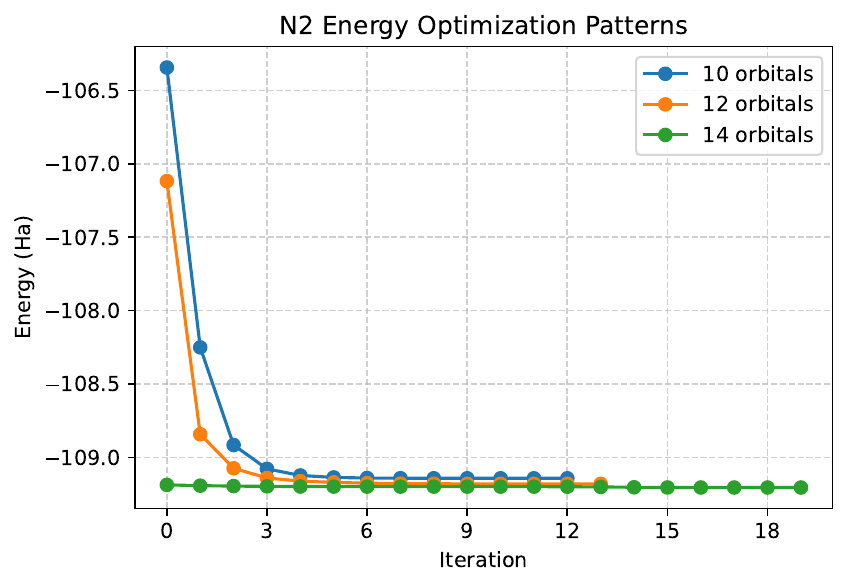}%
    }
    \subfloat[Zoom-in of the \ce{N2} simulation with $14$ orbitals.]{%
        \label{fig:N2_14_convergence}
        \includegraphics[width=0.46\linewidth]{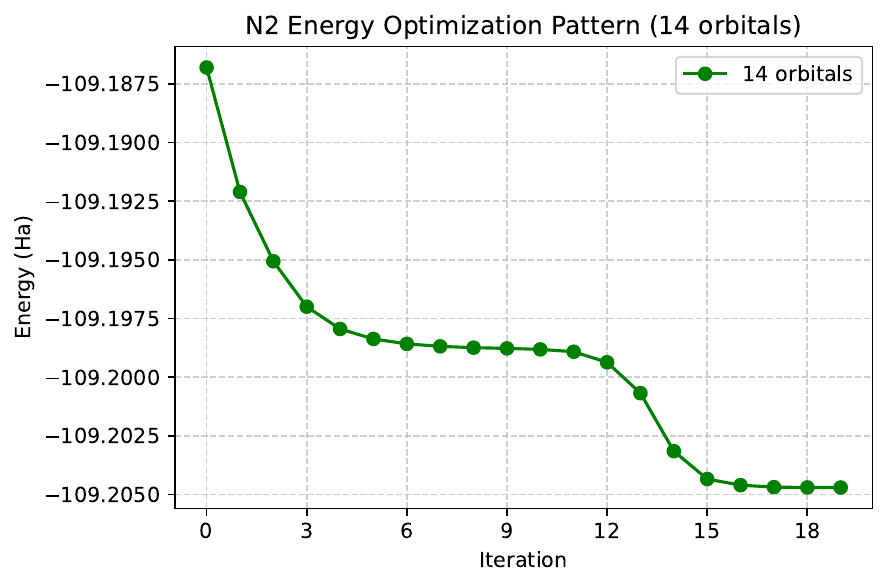}%
    }
    \caption{Convergence patterns of all the considered systems with energy threshold $\delta=10^{-5}\text{ Ha}$. The initial guesses are specified for each system.}
    \label{fig:convergence_plots}
\end{figure*}

In this section, we report the convergence plots for the systems simulated in Section~\ref{sec_numerical_results}.
While experimenting with the algorithm, we have considered different starting basis sets:
\begin{itemize}
    \item Standard atomic orbital basis sets, such as STO-$3$G or cc-pVDZ. Given the separability of the Gaussians defining the orbitals, the initialization is very fast (even if often inaccurate).
    \item Slater-type orbitals. More accurate than the Gaussian basis sets, they are not separable and, therefore, require more time to be initialized in an MRA framework.
    \item Augmented or truncated atomic orbitals basis sets. This often gives a poor starting guess.
    \item Augmented basis sets from previously converged \ac{MRA} orbitals. Good starting guess, it doesn't guarantee a faster convergence.
\end{itemize}

In all cases, the plots in Fig.~\ref{fig:convergence_plots} show a smooth convergence to energy precision $\delta=10^{-5} \text{ Ha}$.

While some cases like \ce{He} and \ce{HeH2}, respectively with $14$ and $6$ orbitals, show a particularly poor initial guess, the optimization still manages to converge in a limited number of iterations.
The \ce{N2} experiment with $14$ orbitals is the most complicated system we have dealt with so far, given the relatively high number of configurations and the molecule's multiconfigurational character.
For this reason, we augmented the previously converged $12$ orbitals basis as a starting guess.

In conclusion, we observe that the MRA+\ac{DMRG} algorithm has only a weak dependence on the starting orbitals, at least for such small systems.

\bibliography{bibliography, DMRG_paper, software}

\end{document}